\documentclass[a4paper,12pt]{article}
\usepackage[british]{babel}
\usepackage{amsfonts}
\usepackage{amsfonts}
\usepackage{amsmath}
\usepackage{amssymb}
\usepackage{graphicx}
\begin{document}

\begin{center}
\baselineskip=1.5
\normalbaselineskip{\Large On discrete stochastic processes with long-lasting time dependence in the variance}

{\large S\'{\i}lvio M. Duarte Queir\'{o}s}
\footnote{Previous address: Centro Brasileiro de Pesquisas F\'{\i}sicas, Rua Dr. Xavier Sigaud 150,
22290-180, Rio de Janeiro-RJ, Brazil \\
email address: Silvio.Queiros@unilever.com, sdqueiro@gmail.com}

\baselineskip=1.0 \normalbaselineskip

\textit{Unilever R\&D Port Sunlight, Quarry Road East, Wirral, CH63 3JW UK \\%
[0pt]
}

\baselineskip=1.0 \normalbaselineskip

{\small (19th September 2008)}
\end{center}

\baselineskip=1.0 \normalbaselineskip

\subsection*{\protect\bigskip Abstract}

In this manuscript, we analytically and numerically study
statistical
properties of an heteroskedastic process based on the celebrated ARCH
generator of random variables whose variance is defined by a memory of
$q_{m}$-exponencial, form ($e_{q_{m}=1}^{x}=e^{x}$). Specifically, we
inspect the
self-correlation function of squared random variables as well as the
kurtosis. In addition, by numerical procedures, we infer the stationary
probability density function of both of the heteroskedastic random
variables and the variance,
the multiscaling properties, the first-passage times distribution, and
the dependence degree.
Finally, we introduce an asymmetric variance version of the model that
enables us to reproduce
the so-called leverage effect in financial markets.

\section{Introduction}

Many of the so-called complex systems are characterised by having time
series with a peculiar feature: although the quantity under measurement
presents an autocorrelation function at noise level for all time lags, when
the autocorrelation of the magnitudes is appraised, a slow and asymptotic
power-law decay is found. This occurs, \textit{e.g.}, with (log) price
fluctuations of several securities traded in financial markets~\cite{livros}%
, temperature fluctuations~\cite{campbell}, neuromuscular activation signals~%
\cite{martin-guerrero} or even fluctuations in presidential approval ratings~%
\cite{gronke} amongst many others~\cite{pier}. Moreover, most of these time
series are also characterised by probability density functions with
asymptotic power-law decay and a profile suggestive of intermittency that is
identified by regions of quasi-laminarity interrupted by spikes. In this
perspective, this type of time series might be seen as a succession of
measurements with a time-dependent standard deviation. Mathematically, this
type of stochastic process is defined as \textit{heteroskedastic}, in
opposition to the class of processes with constant standard deviation that
is defined as homoskedastic. With the primary goal of reproducing and
forecasting inflation time series, it was introduced in $1982$ the
autoregressive conditional heteroskedasticity process ($ARCH$)~\cite{engle}.
The $ARCH$ process rapidly has come to be a landmark in econometrics giving
raise to several generalisations and widespread applications not only in
Economics and Finance but in several other fields as well.

In the sequel of this article, we introduce further insight into a variation
of the $ARCH$ process studied in Ref.~\cite{q-archepl} which is able to
reproduce the properties we have referred to here above. Our considerations
are made both on analytical and numerical grounds. Although the primary goal
of this manuscript is an extensive description of the model following the
lines of Ref.~\cite{q-archepl}, some assessment of its capability of
reproducing the same features of $SP500$ daily log fluctuations spanning the
$3^{rd}$ January $1950$ to the $28^{th}$ February $2007$ is made. In this
context, we also introduce a slight modification on the model which turns it
able to reproduce the leverage effect when the model is applied to surrogate
price fluctuations time series. The manuscript is organised as follows:
after introducing the $ARCH$ processes and present some general properties, we
make known in Sec.~\ref{analytical} some analytical calculations on the
autocorrelation function of the model herein analysed and the correlation
between variables and squared standard deviation for the extension as well as
the expressions for the kurtosis. In Sec.~\ref{numerical}, we introduce
results from the numerical analysis about the probability density functions
of the stochastic variable, $z_{t}$, and its squared instantaneous standard
deviation, $\sigma _{t}^{2}$; The dependence degree between $z ^{t} _{t}$ and
$z ^{2} _{t+\tau }$; The distribution of first passage times of $z_{t}^{2}$;
and the multiscaling properties. In Sec.~\ref{assimetrico}, we establish an
asymmetric variation of the model which allows the reproduction of the
so-called leverage effect. Final considerations are addressed to Sec.~\ref%
{final-remarks}.

\section{The symmetric variance model~\label{analytical}}

We start defining an autoregressive conditional heteroskedastic ($ARCH$)
time series as a discrete stochastic process, $z_{t}$,
\begin{equation}
z_{t}=\sigma _{t}\ \omega _{t},  \label{arch-def}
\end{equation}%
where $\omega _{t}$ is an independent and identically distributed random
variable with mean equal to zero and second-order moment equal to one,
\textit{i.e.}, $\left\langle \omega _{t}\right\rangle =0$ and $\left\langle
\omega _{t}^{2}\right\rangle =1$. Usually, $\omega $ is associated with a
Gaussian distribution, but other distributions of $\omega $ have been
presented to mainly describe price fluctuations~\cite{noise-gen}. In his
benchmark article of Ref.~\cite{engle}, \textsc{Engle} suggested a dynamics
for $\sigma _{t}^{2}$ establishing it as a linear function of past squared
values of $z_{t}$,
\begin{equation}
\sigma _{t}^{2}=a+\sum\limits_{i=1}^{s}b_{i}\ z_{t-i}^{2},\qquad \left(
a,b_{i}\geq 0\right) .  \label{arch-vol}
\end{equation}%
In financial practise, namely price fluctuation modelling, the case $s=1$ ($%
b_{1}\equiv b$) is, by far, the most studied and applied of all $ARCH$-like
processes. It can be easily verified, even for all $s$, that, although $%
\left\langle z_{t}\ z_{t^{\prime }}\right\rangle \sim \delta _{t\,t^{\prime
}}$, correlation $\left\langle \left\vert z_{t}\right\vert \ \left\vert
z_{t^{\prime }}\right\vert \right\rangle $ is not proportional to $\delta
_{t\,t^{\prime }}$. As a matter of fact, for $s=1$, it has been proved that, $%
\left\langle z_{t}^{2}\ z_{t^{\prime }}^{2}\right\rangle $ decays as an
exponential law with a characteristic time $\tau \equiv \left\vert \ln
b\right\vert ^{-1}$, which does not reproduce empirical evidences. In
addition, the introduction of a large value for parameter $s$ bears
implementation problems~\cite{boller}. Expressly, large values of $s$ soar
the complexity of finding the appropriate set of parameters $\left\{
b_{i}\right\} $ for the problem under study as it corresponds to the
evaluation of a large number of fitting parameters. Aiming to solve the
imperfectness of the original $ARCH$ process, the $GARCH\left( s,r\right) $
process was introduced~\cite{granger}, with Eq. (\ref{arch-vol}) being
replaced by,
\begin{equation}
\sigma _{t}^{2}=a+\sum_{i=1}^{s}b_{i}\
z_{t-i}^{2}+\sum\limits_{i=1}^{r}c_{i}\ \sigma _{t-i}^{2}\left(
a,b_{i},c_{i}\geq 0\right) ,  \label{garch}
\end{equation}%
Nonetheless, even this process, presents a exponential decay for $%
\left\langle z_{t}^{2}\ z_{t^{\prime }}^{2}\right\rangle $, with $\tau
\equiv \left\vert \ln \left( b+c\right) \right\vert ^{-1}$ for $GARCH\left(
1,1\right) $, though condition, $b+c<1$, guarantees that $GARCH\left(
1,1\right) $ corresponds exactly to an infinite-order $ARCH$ process~\cite%
{goug}.

Despite the fluctuation of the instantaneous volatility, the $ARCH(1)$
process is actually stationary with a \textit{stationary variance}, given
by,
\begin{equation}
\left\langle \sigma ^{2}\right\rangle =\widehat{\sigma ^{2}}=\frac{a}{1-b}%
,\qquad (b>1),  \label{statvar}
\end{equation}%
($\left\langle \ldots \right\rangle $ represents averages over samples and $%
\widehat{\ldots }$ averages over time). Moreover, it presents a stationary
probability density function (PDF), $P\left( z\right) $, with larger
kurtosis than the kurtosis of distribution $P(\omega )$. The fourth-order
moment is,
\begin{equation*}
\left\langle z^{4}\right\rangle =a^{2}\left\langle \omega ^{4}\right\rangle
\frac{1+b}{\left( 1-b\right) \left( 1-b^{2}\left\langle \sigma
^{4}\right\rangle \right). }
\end{equation*}%
This kurtosis excess is precisely the outcome of time-dependence of $\sigma
_{t}$ . Correspondingly, when $b=0$, the process reduces to generating a
signal with the same PDF of $\omega $, but with a standard variation $\sqrt{a%
}$.

In the remaining of this article we consider a $ARCH\left( 1\right) $
process where an effective immediate past return, $\tilde{z}_{t-1}$, is
assumed in the evaluation of $\sigma _{t}^{2}$ \cite{q-archepl}. Explicitly,
Eq. (\ref{arch-vol}) is replaced by
\begin{equation}
\sigma _{t}^{2}=a+b\,\tilde{z}_{t-1}^{2},\qquad \left( a,b\geq 0\right) ,
\label{vol-qarch}
\end{equation}
where the effective past return is calculated according to
\begin{equation}
\tilde{z}_{t-1}^{2}=\sum\limits _{i=t_{0}}^{t-1}\mathcal{K}\left(
i-t+1\right) \,z_{i}^{2},  \label{zefect}
\end{equation}
where
\begin{equation}
\mathcal{K}\left( t^{\prime }\right) =\frac{1}{\mathcal{Z}_{q_{m}}\left(
t^{\prime }\right) }\exp _{q_{m}}\left[ \frac{t^{\prime }}{T}\right] ,\qquad
\left( t^{\prime }\leq 0,T>0,q_{m}<2\right)  \label{kernel}
\end{equation}
with
\begin{equation}
\exp _{q}\left[ x\right] =e_{q}^{x}\equiv \left[ 1+\left( 1-q\right) \,x%
\right] _{+}^{\frac{1}{1-q}},\qquad \left( q\in \mathcal{R} \right) ,
\label{q-exp}
\end{equation}
$\mathcal{Z}_{q_{m}}\left( t^{\prime }\right) \equiv \sum_{i=-t^{\prime
}}^{0}\exp _{q_{m}}\left[ \frac{i}{T}\right] $ ($\left[ x\right] _{+}=\max
\left\{ 0,x\right\} $ \footnote{%
This condition is known in the literature as \textit{Tsallis cut off} at $%
x=\pm \left( 1-q\right) ^{-1}$.}), known in the literature as $q$\textit{%
-exponential}~\cite{ct-bjp}. This prososal can be enclosed in the
fractionally integrated class of heteroskedastic process ($FIARCH$).
Although it is similar to other proposals~\cite{fiarch}, it has a simpler
structure which permits some analytical considerations without introducing
any underperformance when used for mimicry purposes. For $q=-\infty $, we
obtain the regular $ARCH\left( 1\right) $ and for $q=1$, we have $\mathcal{K}%
\left( t^{\prime }\right) $ with an exponential form since $\exp _{1}\left[ x%
\right]=e^{x}$~\cite{dose}. Although it has a non-normalisable kernel, let
us refer that the value $q_{m}=\infty $ corresponds to the situation that
all past values of $z_{t}$ have the same weight,
\begin{equation}
\mathcal{K}\left( t^{\prime }\right) =\frac{1}{t^{\prime }-t_{0}+1}.
\label{mean-field}
\end{equation}
In this case, memory effects are the strongest possible, \textit{i.e.},
every single element of that past has the same degree of influence on $%
\sigma _{t}^{2}$ making it constant after a few steps. Because of this, in
this case, $P(z)$ is the same as noise $\omega $, as shown in~\cite%
{q-archepl}. Similar heuristic arguments are the base for the Gaussian
nature of the \textquotedblleft Elephant random walk\textquotedblright \cite%
{elephant}.

\medskip

Assuming stationarity in the process some calculations can be made\footnote{%
This has been numerically analysed by computing the $z^{2}_{t}$
self-correlation function for different waiting times}. Namely, it is
provable that the average value of $\sigma ^{2}$ yields,
\begin{equation}
\left\langle \sigma ^{2}\right\rangle =\widehat{\sigma ^{2}}=\frac{a}{1-b}%
,\qquad (b>1),  \label{sigma2-qarch}
\end{equation}
and the covariance, $\left\langle z_{t}\,z_{t^{\prime }}\right\rangle $
corresponds to
\begin{equation}
\left\langle z_{t}\,z_{t^{\prime }}\right\rangle =\left\langle \sigma
_{t}\,\sigma _{t^{\prime }}\right\rangle \left\langle \omega _{t}\,\omega
_{t^{\prime }}\right\rangle ,  \label{covariancez}
\end{equation}
which, due to the uncorrelated nature of $\omega $, gives $\left\langle
z_{t}\,z_{t^{\prime }}\right\rangle =0$ for every $t\neq t^{\prime }$ and $%
\left\langle z_{t}^{2}\right\rangle =\left\langle \sigma ^{2}\right\rangle $%
. In addition, we can verify that all odd moments of $z_{t}$ are equal to
zero. Concerning the fourth-order moment, $\left\langle
z_{t}^{4}\right\rangle $, we have
\begin{equation}
\left\langle z_{t}^{4}\right\rangle =\left\langle \sigma _{t}^{2}\,\sigma
_{t}^{2}\,\omega _{t}^{2}\,\omega _{t}^{2}\right\rangle =\left\langle \left[
\sigma _{t}^{2}\right] ^{2}\right\rangle \left\langle \omega
_{t}^{4}\right\rangle ,  \label{4momento-a}
\end{equation}
which by expansion yields,
\begin{equation}
\begin{array}{cc}
\left\langle z_{t}^{4}\right\rangle = & a^{2}\left\langle \omega
_{t}^{4}\right\rangle +2a\,b\sum\limits _{i=t_{0}}^{t-1}\mathcal{K}\left(
i-t+1\right) \left\langle z_{i}^{2}\right\rangle \left\langle \omega
_{t}^{4}\right\rangle + \\
&  \\
& b^{2}\sum\limits _{i=t_{0}}^{t-1}\left[ \mathcal{K}\left( i-t+1\right) %
\right] ^{2}\left\langle z_{i}^{4}\right\rangle \left\langle \omega
_{t}^{4}\right\rangle + \\
&  \\
& 2b^{2}\sum\limits _{i=t_{0}}^{t-1}\sum\limits _{j=i+1}^{t-1}\mathcal{K}%
\left( i-t+1\right) \mathcal{K}\left( j-t+1\right) \left\langle
z_{i}^{2}\,z_{j}^{2}\right\rangle \left\langle \omega _{t}^{4}\right\rangle .%
\end{array}
\label{4momento-b}
\end{equation}

If $z_{i}$ and $z_{j}$ are assumed as strictly independent, then $%
\left\langle z_{i}^{2}\,z_{j}^{2}\right\rangle =\left\langle
z_{i}^{2}\right\rangle \left\langle z_{j}^{2}\right\rangle =\left\langle
z_{t}^{2}\,\right\rangle ^{2}$. Assuming stationarity we have,
\begin{equation}
\begin{array}{cc}
\left\langle z_{t}^{4}\right\rangle = & a^{2}\left\langle \omega
_{t}^{4}\right\rangle +2a\,b\,\left\langle z_{t}^{2}\right\rangle
\left\langle \omega _{t}^{4}\right\rangle + \\
&  \\
& b^{2}\left\langle z_{t}^{4}\right\rangle \left\langle \omega
_{t}^{4}\right\rangle \sum\limits _{i=t_{0}}^{t-1}\left[ \mathcal{K}\left(
i-t+1\right) \right] ^{2}+ \\
&  \\
& 2b^{2}\left\langle z_{t}^{2}\,\right\rangle ^{2}\left\langle \omega
_{t}^{4}\right\rangle \sum\limits _{i=t_{0}}^{t-1}\sum\limits _{j=i+1}^{t-1}%
\mathcal{K}\left( i-t+1\right) \mathcal{K}\left( j-t+1\right) ,%
\end{array}%
\end{equation}
or
\begin{equation}
\left\langle z_{t}^{4}\right\rangle _{I}=\frac{a^{2}+2a\,b\,\left\langle
z_{t}^{2}\right\rangle +2b^{2}\left\langle z_{t}^{2}\,\right\rangle ^{2}%
\mathcal{Q}_{1}}{1-b^{2}\left\langle \omega _{t}^{4}\right\rangle \mathcal{Q}%
_{2}}\left\langle \omega _{t}^{4}\right\rangle ,
\end{equation}

with $\mathcal{Q}_{1}=\sum\limits _{i=t_{0}}^{t-1}\sum\limits _{j=i+1}^{t-1}%
\mathcal{K}\left( i-t+1\right) \mathcal{K}\left( j-t+1\right) $ and $%
\mathcal{Q}_{2}=\sum\limits _{i=t_{0}}^{t-1}\left[ \mathcal{K}\left(
i-t+1\right) \right] ^{2}$. On the other hand, we have the other limiting
case, $\left\langle z_{i}^{2}\,z_{j}^{2}\right\rangle=\left\langle
z_{t}^{4}\,\right\rangle $. Equation~(\ref{4momento-b}) is then written as
\begin{equation}
\left\langle z_{t}^{4}\right\rangle _{C}=\frac{a^{2}+2a\,b\,\left\langle
z_{t}^{2}\right\rangle }{1-b^{2}\left\langle \omega _{t}^{4}\right\rangle
\left( 2\mathcal{Q}_{1}+\mathcal{Q}_{2}\right) }\left\langle \omega
_{t}^{4}\right\rangle .
\end{equation}
The labelling as upper bound for $\left\langle z_{t}^{4}\right\rangle _{I}$
and lower bound for $\left\langle z_{t}^{4}\right\rangle _{C}$ comes as
follows; The introduction of non-Gaussianity in heteroskedastic processes
comes from the fluctuations in the variance (or in $z_{t}^{2}$), when the
variables are strongly attached between them, there is a small level of
fluctuation in $\sigma _{t}^{2}$ and eventually it becomes constant. With $%
\sigma _{t}$ being a constant, or approximately that, there is not
introduction of a significant level of non-Gaussianity measured from $%
\left\langle z_{t}^{4}\right\rangle $, hence $\left\langle
z_{t}^{4}\right\rangle _{I}>\left\langle z_{t}^{4}\right\rangle _{C}$ ($a
\ne 0$).

\begin{figure}[h]
\centering
\includegraphics[angle=0,width=0.75\columnwidth]{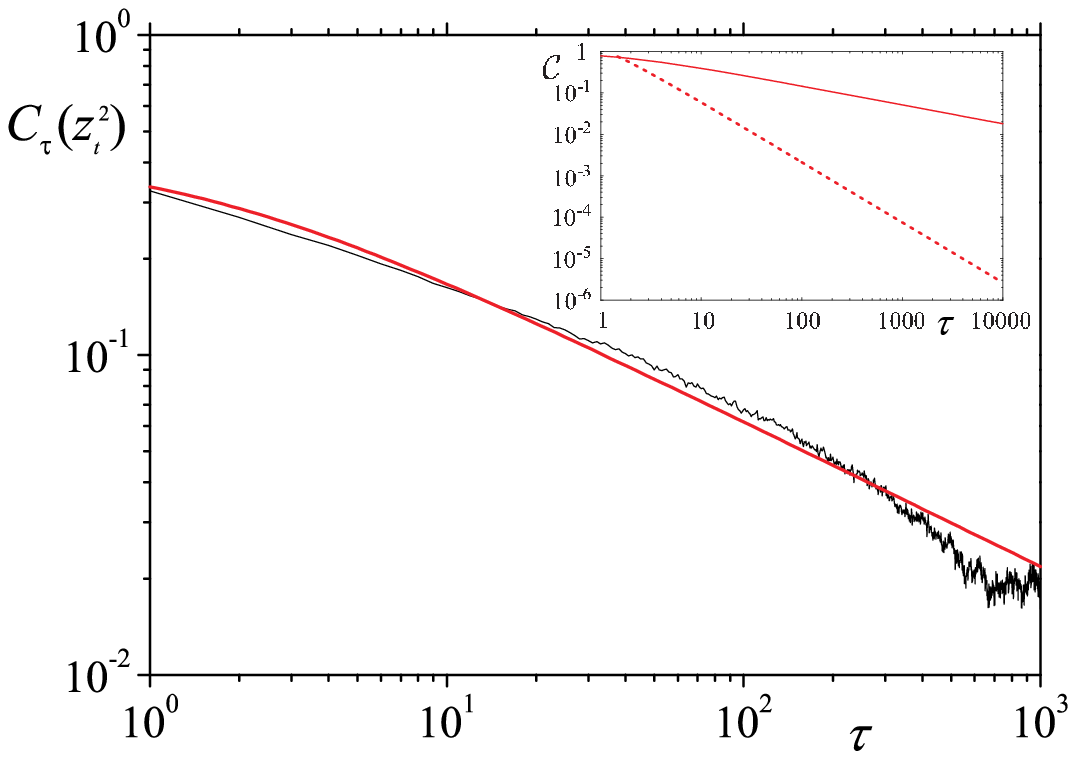}
\caption{The black line presents the numerical evaluation of the $z_{t}^{2}$
autocorrelation function \textit{vs} the lag $\protect\tau $ and the red
line $\mathcal{C}_{2}\left( \protect\tau \right) $ from Eq.~(\protect\ref{c2}%
) for parameters $\left\{ 0.5,b=b_{SP},q_{m}=q_{SP}\right\} $. The inset
depicts the way $\mathcal{C}_{1}\left( \protect\tau \right) $ (dotted line)
and $\mathcal{C}_{2}\left( \protect\tau \right) $ (full line) decay. As it
can be seen, $\mathcal{C}_{1}\left( \protect\tau \right) $ decays much
faster than $\mathcal{C}_{2}\left( \protect\tau \right) $ which is the major
responsible for $C_{\protect\tau }\left( z_{t}^{2}\right) $ behaviour for
large $\protect\tau $.}
\label{fig-corsp500}
\end{figure}

For an accurate description of $\left\langle z_{t}^{4}\right\rangle $, which
lies between the two limiting expressions, we must compute correlations $%
\left\langle z_{t}^{2}\,z_{t^{\prime }}^{2}\right\rangle $. That is obtained
averaging,
\begin{equation}
\begin{array}{cc}
z_{t}^{2}\,z_{t^{\prime }}^{2}= & a^{2}\omega _{t}^{2}\,\omega _{t^{\prime
}}^{2}+a\,b\sum\limits _{i=t_{0}}^{t-1}\mathcal{K}\left( i-t+1\right)
\,z_{i}^{2}\,\omega _{t}^{2}\,\omega _{t^{\prime }}^{2}+ \\
&  \\
& a\,b\sum\limits _{j=t_{0}}^{t^{\prime }-1}\mathcal{K}\left( j-t^{\prime
}+1\right) \,z_{j}^{2}\,\omega _{t}^{2}\,\omega _{t^{\prime }}^{2}+ \\
&  \\
& b^{2}\sum\limits _{i=t_{0}}^{t-1}\sum\limits _{j=t_{0}}^{t^{\prime }-1}%
\mathcal{K}\left( i-t+1\right) \mathcal{K}\left( j-t^{\prime }+1\right)
\,z_{i}^{2}\,z_{j}^{2}\,\omega _{t}^{2}\,\omega _{t^{\prime }}^{2}.%
\end{array}
\label{correlacao1}
\end{equation}
Defining $\tau \equiv t-t^{\prime }$, the last term of rhs of Eq.~(\ref%
{correlacao1}), hereon labelled as $\mathcal{C}$, is responsible for the dependence of $%
\left\langle z_{t}^{2}\,z_{t^{\prime }}^{2}\right\rangle $ with $\tau $. It
can be written as
\begin{equation}
\begin{array}{ccc}
\mathcal{C} & = & b^{2}\omega _{t}^{2}\,\omega _{t^{\prime }}^{2}\sum\limits
_{i=t_{0}}^{t-1}\,\sum\limits _{j=t_{0}}^{t-\tau -1}\mathcal{K}\left(
i-t+1\right) \mathcal{K}\left( j-t+1+\tau \right) \,z_{i}^{2}\,z_{j}^{2} \\
&  &  \\
& \sim & \sum\limits _{i=t_{0}}^{t-\tau -1}\mathcal{K}\left( i-t+1\right)
\mathcal{K}\left( i-t+1+\tau \right) \,z_{i}^{4}+ \\
&  &  \\
&  & \sum\limits _{i=t_{0}}^{t-\tau -1}\sum\limits _{j=t-\tau +1}^{t-1}%
\mathcal{K}\left( i-t+1+\tau \right) \,\mathcal{K}\left( j-t+1\right)
\,z_{i}^{2}\,z_{j}^{2}.%
\end{array}
\label{correlacao2}
\end{equation}
We shall now consider a continuous approximation where the summations are
changed by integrals,
\begin{equation*}
\sum_{i=t_{0}}^{t-1}\ldots \rightarrow \int_{0}^{t}\ldots dx,
\end{equation*}
and $t\gg 1$, so that the following relations are obtained in the limit $%
t\rightarrow \infty $. Computing $\left\langle \mathcal{C}\right\rangle $,
the term in $z_{i}^{4}$ has the coefficient,
\begin{equation}
\begin{array}{ccc}
\mathcal{C}_{1}\left( \tau \right) & \sim & \int_{0}^{t-\tau }\exp _{q_{m}}
\left[ x-t\right] \exp _{q_{m}}\left[ x-t+\tau \right] \,dx \\
&  &  \\
& \underset{t\rightarrow \infty }{\ =} & \frac{\left[ \tau \left(
q_{m}-1\right) \right] ^{\frac{1}{1-q_{m}}}}{2-q_{m}}F_{2,1}\left[ \frac{%
2-q_{m}}{q_{m}-1},\frac{1}{q_{m}-1};2+\frac{1}{q_{m}-1};\frac{1}{\left(
1-q_{m}\right) \tau }\right] ,%
\end{array}
\label{c1}
\end{equation}
where $F_{2,1}\left[ a,b;c;x\right] $ is the generalised hypergeometric
function. Regarding the term in $z_{i}^{2}\,z_{j}^{2}$, the first
approximation is obtained considering $\left\langle
z_{i}^{2}\,z_{j}^{2}\right\rangle =\left\langle z_{i}^{2}\right\rangle
\left\langle z_{j}^{2}\right\rangle =\left\langle z_{t}^{2}\right\rangle
^{2} $. Its coefficient is then given by
\begin{equation}
\begin{array}{ccc}
\mathcal{C}_{2}\left( \tau \right) & \sim & \int_{t-\tau
}^{t}\int_{0}^{t-\tau }\exp _{q_{m}}\left[ y-t\right] \exp _{q_{m}}\left[
x-t+\tau \right] \,dx\,dy \\
&  &  \\
& \underset{t\rightarrow \infty }{\sim } & \exp _{q_{c}}\left[ -\lambda
\,\tau \right] ,%
\end{array}
\label{c2}
\end{equation}
with
\begin{equation}
q_{c}=\frac{1}{2-q_{m}},  \label{qc-qm}
\end{equation}
and $\lambda =q_{c}^{-1}$. A simple inspection shows that $\mathcal{C}
_{2}\left( \tau \right) $ decays much slower than $\mathcal{C}_{1}\left(
\tau \right) $, hence the asymptotic form of $\left\langle
z_{t}^{2}\,z_{t^{\prime }}^{2}\right\rangle $ is dominated by $\mathcal{C}
_{2}\left( \tau \right) $ as it is illustrated in the inset of Fig.~\ref%
{fig-corsp500}. In Fig. \ref{fig-corsp500}, we bring face to face Eq.~(\ref%
{c2}) and the autocorrelation function of $z_{t}^{2}$ from numerical
simulation using the parameters applied to reproduce $SP500$ returns
previously determined, namely $\left\{ a,b=b_{SP},q_{m}=q_{SP}\right\} =
\left\{ 0.5,0.99635,1.6875\right\} $.

\medskip

From all these equations we are able to conjecture expressions which relate
parameters $\left\{ a,b,q_{m}\right\} $ with the form of the distribution
for the case where $\left\langle z_{t}^{4}\right\rangle $ is finite. In this
way, we can use the ansatz that the distribution of this dynamical model is
associated with a $q$-Gaussian (or Student-$t$) distribution \footnote{%
A $q$-Gaussian, with $q>1$, corresponds to a Student-$t$ with $m$ degrees of
freedom with $q=\frac{3+m}{1+m}$ where $m$ is taken as a real positive
number.},
\begin{equation}
p\left( z\right) =\mathcal{A\,}\exp _{q}\left[ -\mathcal{B}\,z^{2}\right]
,\qquad \left( q<3\right) ,  \label{q-gaussian}
\end{equation}%
with $\mathcal{B}=\left[ \bar{\sigma}_{q}^{2}\left( 3-q\right) \right] ^{-1}$%
, where,
\begin{equation*}
\bar{\sigma}_{q}^{2}\equiv \int z^{2}\left[ p\left( z\right) \right]
^{q}dz/\int \left[ p\left( z\right) \right] ^{q}dz,
\end{equation*}%
is the $q$-generalised second order moment~\cite{ct-bjp}, and $\mathcal{A}$
is the normalisation constant. This assumption is based on the same type of
arguments used in~\cite{garch} and whose accuracy we verify later on (see Sec.~%
\ref{numerical}). For $q<5/3$, $\bar{\sigma}_{q}^{2}$ relates to the usual
variance according to
\mbox{$\bar{\sigma}_{q}^{2}\left(
3-q\right) =\bar{\sigma}^{2}\left( 5-3\,q\right) $}.

From Eqs. (\ref{4momento-b}), (\ref{correlacao1}), (\ref{c1}), and (\ref{c2}%
) we can write,
\begin{equation}
\begin{array}{cc}
\left\langle z_{t}^{4}\right\rangle = & a^{2}\left\langle \omega
_{t}^{4}\right\rangle +2a\,b\,\left\langle z_{t}^{2}\right\rangle
\left\langle \omega _{t}^{4}\right\rangle +b^{2}\left\langle
z_{t}^{4}\right\rangle \left\langle \omega _{t}^{4}\right\rangle \frac{%
\left( 2-q_{m}\right) ^{2}}{3-q_{m}} \\
&  \\
& +b^{2}\left\langle \omega _{t}^{4}\right\rangle \left[ a^{2}\left\langle
\omega _{t}^{2}\right\rangle ^{2}+2\,a\,b\,\left\langle
z_{t}^{2}\right\rangle \,\left\langle \omega _{t}^{2}\right\rangle ^{2}+2%
\mathcal{K}\right] .%
\end{array}
\label{z4fit}
\end{equation}
where $\mathcal{K}$ represents terms like,%
\begin{equation}
\int_{0}^{t}\int_{0}^{t}\exp _{q_{m}}\left[ x-t\right] \exp _{q_{m}}\left[
x+\tau -t\right] \left[ \mathcal{C}_{1}\left( \tau \right) +\mathcal{C}%
_{2}\left( \tau \right) \right] \,d\tau \,dx,
\end{equation}%
which corresponds to a quite complex integration over $x$ of hypergeometric
function special functions like $F_{2,1}\left[ \tilde{a},\tilde{b};\tilde{c}%
;x\right] $ and the Appell hypergeometric function~\cite{appell} where $%
\tilde{a}$, $\tilde{b}$, and $\tilde{c}$ represent general values.

Taking into attention that for a $q$-Gaussian with $q<\frac{7}{5}$ its
fourth moment is,
\begin{equation}
\left\langle z^{4}\right\rangle =3\left( \bar{\sigma}^{2}\right) ^{2}\frac{%
3\,q-5}{5\,q-7},  \label{4qgauss}
\end{equation}
we can obtain approximate relations between the parameters of the model and
the parameters of the distribution. This is achieved when we equalise Eqs.~(%
\ref{z4fit}) and (\ref{4qgauss}), remembering the expression of the variance,
Eq. (\ref{sigma2-qarch}), and the form of the autocorrelation function of $%
z_{t}^{2}$. This procedure is obviously important in parameter estimation.
Therefore, from the decay of the $z_{t}^{2}$ autocorrelation function we can
determine the value of $q_{m}$, and $a$ and $b$ from the equalisation we
have just referred to together with Eq.~(\ref{sigma2-qarch}).

\section{Numerical considerations~\label{numerical}}

\subsection{Stationary probability density functions}

\begin{figure}[h]
\centering
\includegraphics[angle=0,width=0.45\columnwidth]{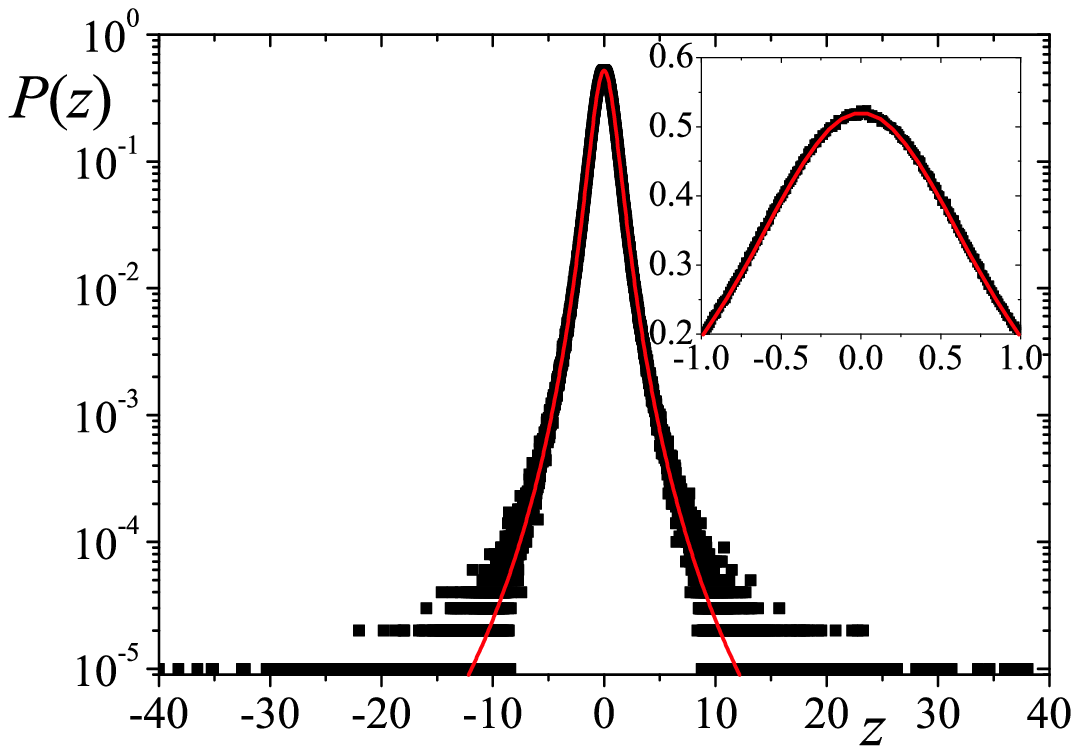} %
\includegraphics[angle=0,width=0.45\columnwidth]{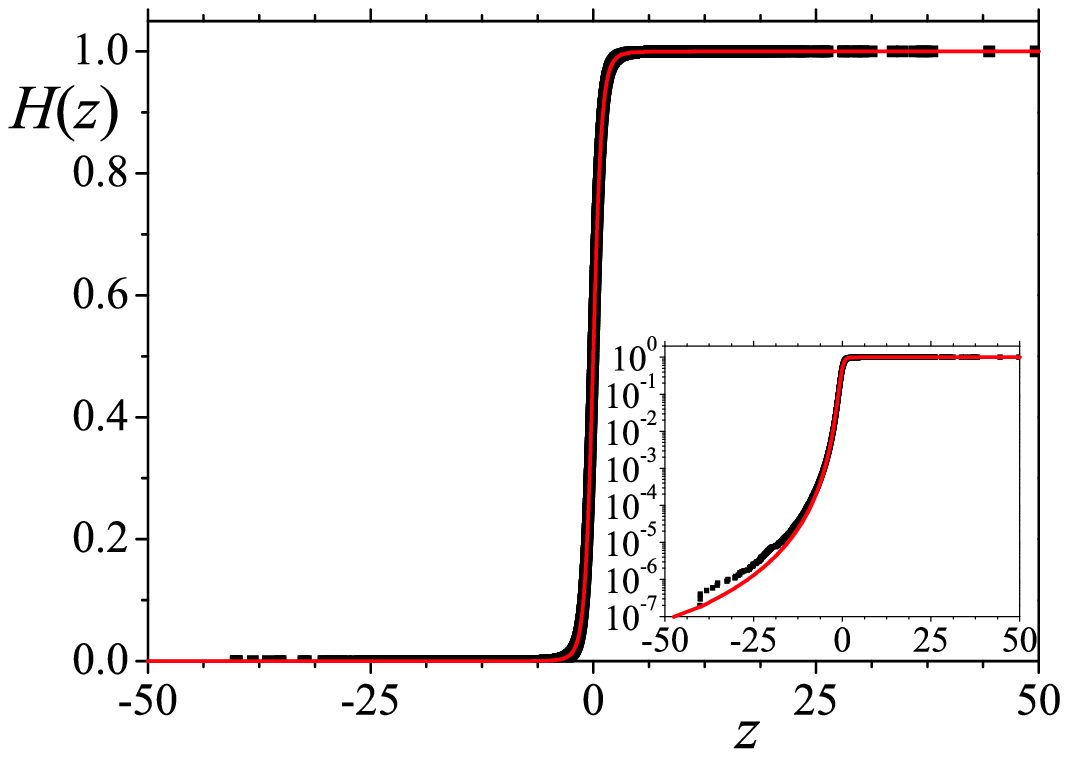}
\caption{Upper panel: The symbols represent the probability density function
$P\left( z\right) $ \textit{vs} $z$ for parameters $q_{m}=1.5$ and $b=0.875$%
, and the line the best fit for a $q$-Gaussian with $q=1.385$ [on log-linear
scale] and unitary standard deviation ($\protect\chi ^{2}=1.39\times 10^{-6}$
and $R^{2}=0.99987$). The inset is the same, but on linear-linear scale
permitting the appraisal of the fitting in the central region. Lower panel:
The symbols are for the empirical cumulative distribution function $H\left(
z\right) $ \textit{vs} $z$ for the same parameters and the line the CDF of a
$q$-Gaussian with $q=1.385$ and unitary standard deviation ($D_{KS}=0.00457$%
). The inset is the same plot, but on a log-linear scale.}
\label{fig-pdf}
\end{figure}

We firstly recover the results previously presented for the adjustment of $%
P\left( z\right) $ with $q$-Gaussians. As mentioned in~\cite{q-archepl} the
adjustment is striking (diagrams of $q$ as a function of $b$ and $q_{m}$ are
presented in Figure 1 of that reference). Using the method of $\chi ^{2}$
minimisation, we have obtained for the same cases previously studied%
\footnote{%
Comparing with prior studies we have increased the runs by a factor of $10$.}
average values of $\chi ^{2}=1.1\times 10^{-6}$ (per degree of freedom) and $%
R^{2}=0.99990$. In Fig.~\ref{fig-pdf}, we present an example for which it is
possible to assent the accuracy of the fitting not only in the tails, but in
the peak of the PDF as well. We have performed further analysis using the
cumulative distribution function (CDF) and the Kolmogorov-Smirnov Distance, $%
D_{KS}$,
\begin{equation}
D_{KS}=\max \left\vert H\left( z\right) -H_{0}\left( z\right) \right\vert ,
\end{equation}
where $H\left( z\right) $ is the empirical CDF obtained from numerical
evaluation of the model and $H_{0}\left( z\right) $ is the testing
probability density function,
\begin{equation}
H_{0}\left( z\right) =\int_{-\infty }^{z}\mathcal{A\,}\exp _{q}\left[ -%
\mathcal{B}\,x^{2}\right] \,dx.
\end{equation}
The average $D_{KS}$ value obtained for the same cases plotted in Figure 1
of the prior work is equal to $4.25\times 10^{-3}$. Such values allow us to
rely on the null hypothesis~\cite{boes},
\begin{equation}
P\left( z\right) =p\left( z\right) =\mathcal{A\,}\exp _{q}\left[ -\mathcal{%
B\,}z^{2}\right] .  \label{nullhypo}
\end{equation}

\begin{figure}[h]
\centering\includegraphics[angle=0,width=0.75\columnwidth]{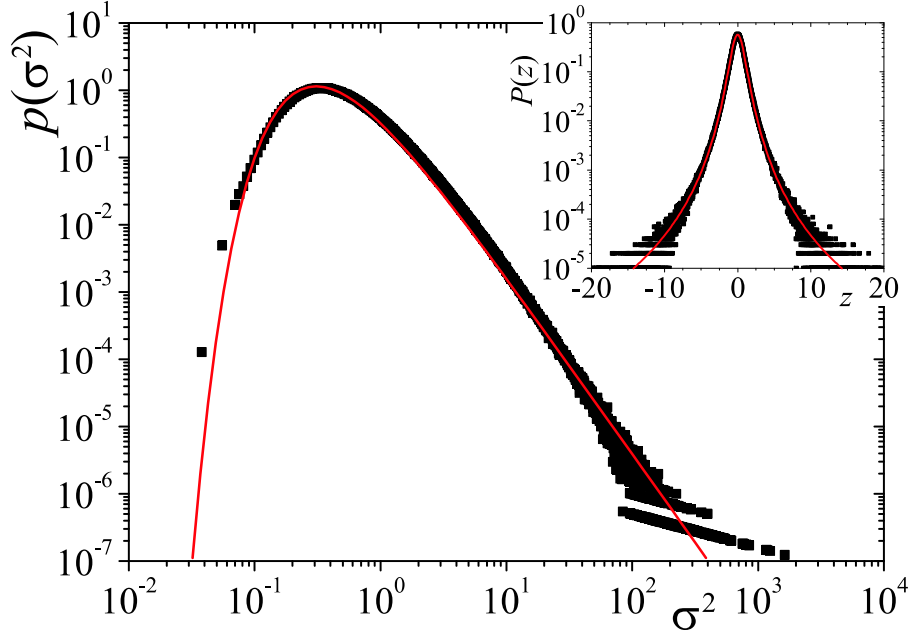}
\caption{The black symbols represent $p_{\protect\sigma }\left( \protect%
\sigma ^{2}\right) $ \textit{vs} $\protect\sigma ^{2}$ [on log-log scale]
obtained by numerical evaluation of the process with $q_{m}=q_{SP}$ and $%
b=b_{SP}$ yielding a $q$-Gaussian distribution with $q=1.465$ ($\protect\chi %
^{2}=1.6\times 10^{-6}$) shown in the inset [on log-linear scale]. The red
line in the main plot is the representation of the inverted Gamma
distribution with $c=1.648\ldots $ and $\protect\theta =0.770\ldots $ ($%
\protect\chi ^{2}=6.1\times 10^{-4}$).}
\label{fig-psigma}
\end{figure}

Based on the acceptance of the null hypothesis (\ref{nullhypo}) we are able
to introduce some insight into the distribution of $\sigma ^{2}$, $p_{\sigma
}\left( \sigma ^{2}\right) $. Firstly, we carry out the following change of
variables,
\begin{equation}
\left\{
\begin{array}{c}
\breve{z}_{t}=\ln \,z_{t}^{2} \\
\breve{\sigma}_{t}=\ln \,\sigma _{t}^{2} \\
\breve{\omega}_{t}=\ln \,\omega _{t}^{2}%
\end{array}%
\right. ,
\end{equation}%
so that Eq.~(\ref{arch-def}) turns into,
\begin{equation}
\breve{z}_{t}=\breve{\sigma}_{t}+\breve{\omega}_{t}.
\end{equation}%
In probability space, regarding that $\breve{\sigma}_{t}$ and $\breve{\omega}%
_{t}$ are independent, we have,
\begin{equation}
p\left( \breve{z}\right) =\int_{-\infty }^{\infty }P_{\breve{\sigma}}\left(
\breve{\sigma}\right) \,P_{\breve{\omega}}\left( \breve{z}-\breve{\sigma}%
\right) \,d\breve{\sigma}.  \label{convolution}
\end{equation}%
We can now apply the convolution theorem. Being $p\left( \breve{z}%
_{t}\right) $, the probability of $\breve{z}_{t}$, according to such a
theorem,
\begin{equation}
p\left( \breve{z}_{t}\right) =\mathcal{F}^{-1}\left[ \check{P}_{\breve{\sigma%
}}\left( k\right) \check{P}_{\breve{\omega}}\left( k\right) \right] ,
\label{invfourier}
\end{equation}%
where
\begin{equation}
\check{P}_{x}\left( k\right) =\frac{1}{\sqrt{2\,\pi }}\int f\left( x\right)
\exp \left[ i\,k\,x\right] \,dx\equiv \mathcal{F}\left[ f\left( x\right) %
\right] ,
\end{equation}%
and $\mathcal{F}^{-1}\left[ f_{x}\left( k\right) \right] =$ $\frac{1}{\sqrt{%
2\,\pi }}\int f_{x}\left( k\right) \exp \left[ -i\,k\,x\right] \,dx=f\left(
x\right) $ is the inverse Fourier Transform. Since we respectively know and
postulate the form of $p\left( \omega \right) $ and $p\left( z\right) $, we
can write down,
\begin{equation}
\begin{array}{c}
p\left( \breve{\omega}\right) =\frac{1}{\sqrt{2\,\pi }}\exp \left[ \frac{%
\breve{\omega}}{2}-\frac{e^{\breve{\omega}}}{2}\right] , \\
\\
p\left( \breve{z}\right) =\mathcal{A\,}\exp _{q}\left[ -\mathcal{B}\,e^{\,%
\breve{z}}\right] \exp \left[ \frac{\breve{z}}{2}\right] ,%
\end{array}%
\end{equation}%
($\mathcal{B=B}\left( \bar{\sigma}=1\right) $), yielding the respective
Fourier Transforms \cite{fouriertransforms},%
\begin{equation}
\begin{array}{c}
\check{P}_{\breve{\omega}}\left( k\right) =\frac{2^{i\,k-\frac{1}{2}}}{\pi }%
\Gamma \left[ \frac{1}{2}+i\,k\right] , \\
\\
\begin{array}{cc}
\check{P}_{\breve{z}}\left( k\right) = & \frac{1}{\sqrt{2\pi }}\mathcal{A}%
\left\{ \left( -1\right) ^{-Q}\left( \mathcal{B\,}q-\mathcal{B}\right) ^{%
\frac{1}{1-q}-Q}B\left[ \frac{1}{\mathcal{B}-\mathcal{B\,}q},Q,\frac{2-q}{q-1%
}\right] \right. + \\
&  \\
& \left. \Gamma \left[ \frac{1}{2}+i\,k\right] \tilde{F}_{2,1}\left[ \frac{1%
}{q-1},\frac{1}{2}+ik,\frac{3}{2}+ik,\mathcal{B}-\mathcal{B\,}q\right]
\right\} .%
\end{array}%
\end{array}
\label{invfourierp}
\end{equation}%
where $Q=\frac{1}{1-q}+\frac{1}{2}+ik$, $B\left[ \ldots \right] $ is the
Beta function, and $\tilde{F}_{2,1}\left[ \ldots \right] $ is the
regularised hypergeometric function~\cite{appell}. Applying Eq.~(\ref%
{invfourierp}) in Eq. (\ref{invfourier}) we can compute the distribution of $%
\breve{\sigma}$ (easily related to $p_{\sigma }\left( \sigma \right) $),
\begin{equation}
P_{\breve{\sigma}}\left( \breve{\sigma}\right) =\mathcal{F}^{-1}\left[ \frac{%
\check{P}_{\breve{z}}\left( k\right) }{\check{P}_{\breve{\omega}}\left(
k\right) }\right] ,  \label{invfouriervol}
\end{equation}%
From a laborious and tricky calculation, using properties of $\tilde{F}_{2,1}%
\left[ \ldots \right] $ (see \cite{functions} and related properties), it can be verified
$P_{\breve{\sigma}}\left( \breve{\sigma}\right) $ cooresponds to,
\begin{equation}
\check{P}_{\breve{\sigma}}\left( k\right) =\frac{\Gamma \left[ \theta -i\,k%
\right] }{\left( 2\beta \right) ^{i\,k}\Gamma \left[ c\right] }.
\end{equation}
Therefore, $\sigma ^{2}$ follows an inverse Gamma distribution,
\begin{equation}
p_{\sigma }\left( \sigma ^{2}\right) =\frac{1}{\left( 2\,\theta \right)
^{c}\,\Gamma \left[ c\right] }\left( \sigma ^{2}\right) ^{-1-c}\exp \left[ -%
\frac{1}{2\,\theta \,\sigma ^{2}}\right] ,  \label{invgamma}
\end{equation}
where $c=\frac{3-q}{2\,q-2}$ and $\theta =\frac{q-1}{\bar{\sigma}^{2}\left(
5-3\,q\right) }$. This result attests the validity of the superstatistical
approach to the problem of heteroskedasticity. It is worth mentioning that
superstatistics \cite{beck-cohen} represents the long-term statistics in
systems with fluctuations in some characteristic intensive parameter of the
problem like the dissipation rate in Lagragian turbulent fluids \cite{reynolds} or the
standard deviation like in the subject matter of heteroskedasticity. For the values
of $q_{SP}$ and $b_{SP}$, with have obtained random variables $z_{t}$
associated with a $\left( q=1.465\right) $-Gaussian. This yields $%
c_{SP}=1.648\ldots $ and $\theta _{SP}=0.770\ldots $, which have been
applied in Fig. \ref{fig-psigma} to fit $p_{\sigma }\left( \sigma
^{2}\right) $ obtained by numerical procedures. In that plot, it is visible
that numerical and analytical curves are in proximity.

\subsection{Dependence degree}

The degree that the elements of a time series are tied-in is not completely
expressed by the correlation function in the majority of the cases. In fact,
regarding its intimate relation with the covariance, the correlation
function is only a measure of linear dependences. Aiming to assess
non-linear dependences, information measures have been widely applied~\cite%
{physrepaus}. In our case, we use a non-extensive generalisation of
Kullback-Leibler information measure~\cite{ct-kl,ct-plastino-borland}%
\footnote{%
In the limit $q\rightarrow 1$ the Kullback-Leibler mutual information
definition is recovered.},
\begin{equation}
I_{q}\equiv -\sum_{t}p\left( z_{t}^{2},z_{t+\tau }^{2}\right) \frac{\left[
\frac{p^{\prime }\left( z_{t}^{2},z_{t+\tau }^{2}\right) }{p\left(
z_{t}^{2},z_{t+\tau }^{2}\right) }\right] ^{1-q}-1}{1-q},  \label{qKL}
\end{equation}%
where $p^{\prime }\left( z_{t}^{2},z_{t+\tau }^{2}\right) =p\left(
z_{t}^{2}\right) p\left( z_{t+\tau }^{2}\right) =\left[ p\left( z^{2}\right) %
\right] ^{2}$ (assuming stationarity), which has been able to provide a set
of interesting results with respect to dependence problems~\cite%
{kl-utilizacao}. The quantification of the dependence degree is made through
a value, $q^{op}$, which corresponds to the inflexion point of the
normalised version of $I_{q}$,
\begin{equation}
R_{q}\equiv \frac{I_{q}}{I_{q}^{\max }},  \label{rq}
\end{equation}%
where $I_{q}^{\max }$ is the value of $I_{q}$ when variables $z\left(
t\right) $ and $z\left( t+\tau \right) $ present a biunivocal dependence
(see full expression in Ref.~\cite{ct-plastino-borland}).

For infinite signals it can be shown that, when the variables are completely
independent $q^{op}=\infty $, whereas $q^{op}=0$ when variables are
one-to-one dependent. For finite systems, there is a noise level, $%
q_{n}^{op} $, which is achieved after a finite time lag $\tau $. Typical
curves of $R_{q}$ are depicted in Fig.~\ref{fig-rq} for $q_{m}=q_{SP}$ and $%
b=b_{SP}$.

\begin{figure}[h]
\centering\includegraphics[angle=0,width=0.75\columnwidth]{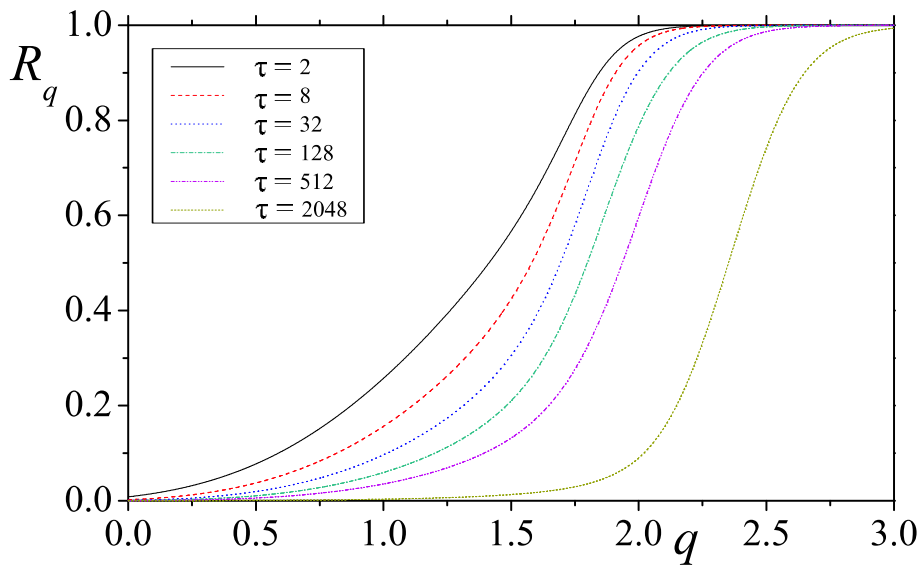}
\caption{Normalised generalised mutual entropy $I_{q}$ \textit{vs} $q$ for
parameters $\left\{ q_{m}=q_{SP},b=b_{SP}\right\} $ and values of the lag
presented in the figure.}
\label{fig-qop}
\end{figure}

\begin{figure}[h]
\centering\includegraphics[angle=0,width=0.75\columnwidth]{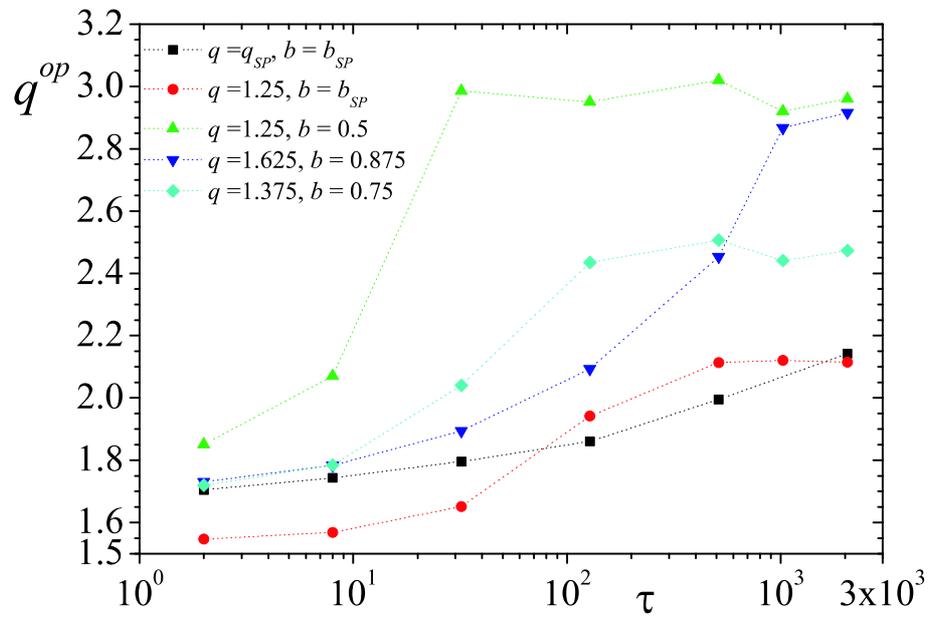}
\caption{Values of $q^{op}$ \textit{vs} time lag $\protect\tau $ for the
pairs $\left\{ q_{m}=q_{SP},b=b_{SP}\right\} $ as indicated in the figure.
The dotted lines are merely presented as a guide to the eye. }
\label{fig-rq}
\end{figure}

In what follows, we present results obtained from numerical evaluation of $%
R_{q}$ for different values of $\tau $. As expected, dependence relies on
the balance between the extension of memory, which is given by $q_{m}$ and
the weight of effective past value, $\tilde{z}_{t-1}$, on $\sigma _{t}$.
Firstly, let us compare cases $\left\{ q_{m}=q_{SP},b=b_{SP}\right\} $ and $%
\left\{ q_{m}=1.25,b=b_{SP}\right\} $, as an example of what happens when we
fix $b$ as a constant (see Fig.~\ref{fig-qop}). Dependence is obviously
long-lasting in the former case, in the sense that it takes longer to attain
$q_{n}^{op}$, but for small values of $\tau $, the latter has presented
higher levels of dependence. This has to do that $\mathcal{K}%
\left( t^{\prime }\right) $ is normalised and that implies the intersection
of the curves for different values of $q_{m}$ at some value of $t^{\prime }$%
. Alternatively, when $q_{m}$ decreases, the recent values of $z_{t}$ have
more influence on $\tilde{z}_{t-1}$ than past values. When the value of $%
q_{m}$ is kept constant, we have verified that smaller values of $b$ lead to
a faster approach to noise value $q_{n}^{op}$. In a previous work on $%
GARCH\left( 1,1\right) ~$~\cite{garch}, we have verified that variables
approximately associated with the same distribution present the same level
of dependence independently of the pair $\left( b,c\right) $ chosen. In
this case, recurring to cases $\left\{ q_{m}=1.375,b=0.75\right\} $ and $%
\left\{ q_{m}=1.625,b=0.875\right\} $, we have noticed that the curves
present very close values for small lags, but they fall apart for $\tau >10$,
revealing a more intricate relation between $q$, $q_{m}$, and $b$ than in $%
GARCH\left( 1,1\right) $. Additionally, comparing dependence and correlation,
we have verified that the decay is faster for the latter. Specifically,
taking into account noise values of $q^{op}$ and $C_{\tau }\left(
z_{t}^{2}\right) $, it is verifiable that $q^{op}$ takes longer to
achieve $q_{n}^{op}$ than $C_{\tau }\left( z_{t}^{2}\right) $ takes to reach
$C_{n}\left( z_{t}^{2}\right) $.

\subsection{First-Passage Times}

First-passage studies in stochastic processes are of considerable interest.
Not only from the scientific point of view~\cite{risken} (it is useful in
the approximate calculation of the lifetimes of the problems/systems) as
well as from a practical perspective, since they can be applied to
quantify the extent of reliability of forecasting procedures, \textit{e.g.},
in meteorology or finance~\cite{livros,stanley-returns,climate}. In what it
is next to come, we have analysed the probability of $z_{t}^{2}\in S_{i}=%
\left[ a,b\right) $ and $z_{t+T}^{2}\in S_{i}$. We have divided $z_{t}^{2}$
domain into five different intervals. Explicitly:

\begin{itemize}
\item $S_{1}$: $z_{t}^{2}\leq 1$;

\item $S_{2}$: $1<$ $z_{t}^{2}\leq 2$;

\item $S_{3}$: $2<$ $z_{t}^{2}\leq 5$;

\item $S_{4}$: $5<$ $z_{t}^{2}\leq 10$;

\item $S_{5}$: $z_{t}^{2}>10$.
\end{itemize}

\begin{figure}[h]
\centering
\includegraphics[angle=0,width=0.45\columnwidth]{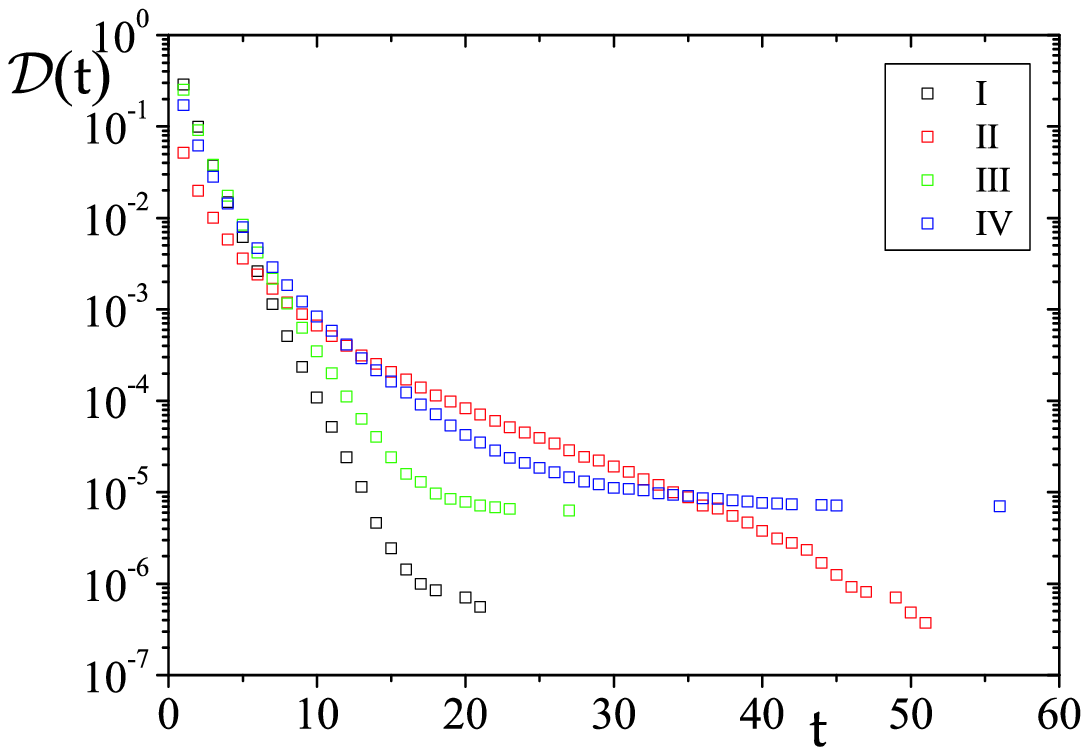} %
\includegraphics[angle=0,width=0.45\columnwidth]{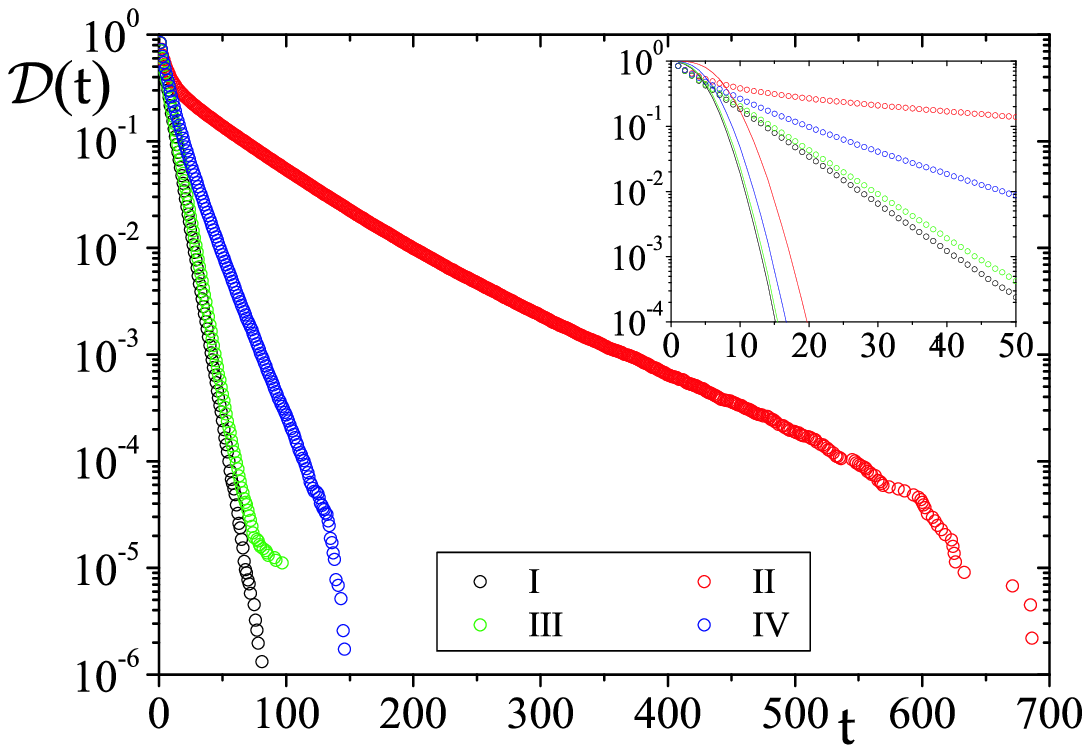} %
\includegraphics[angle=0,width=0.45\columnwidth]{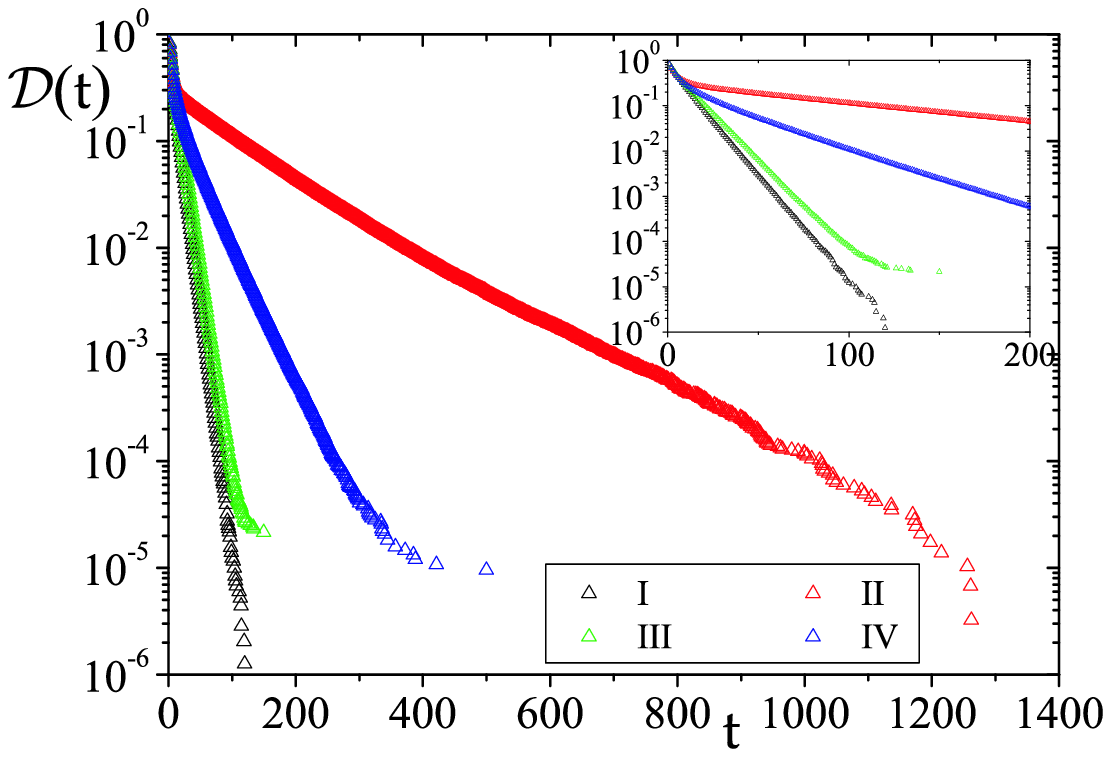} %
\includegraphics[angle=0,width=0.45\columnwidth]{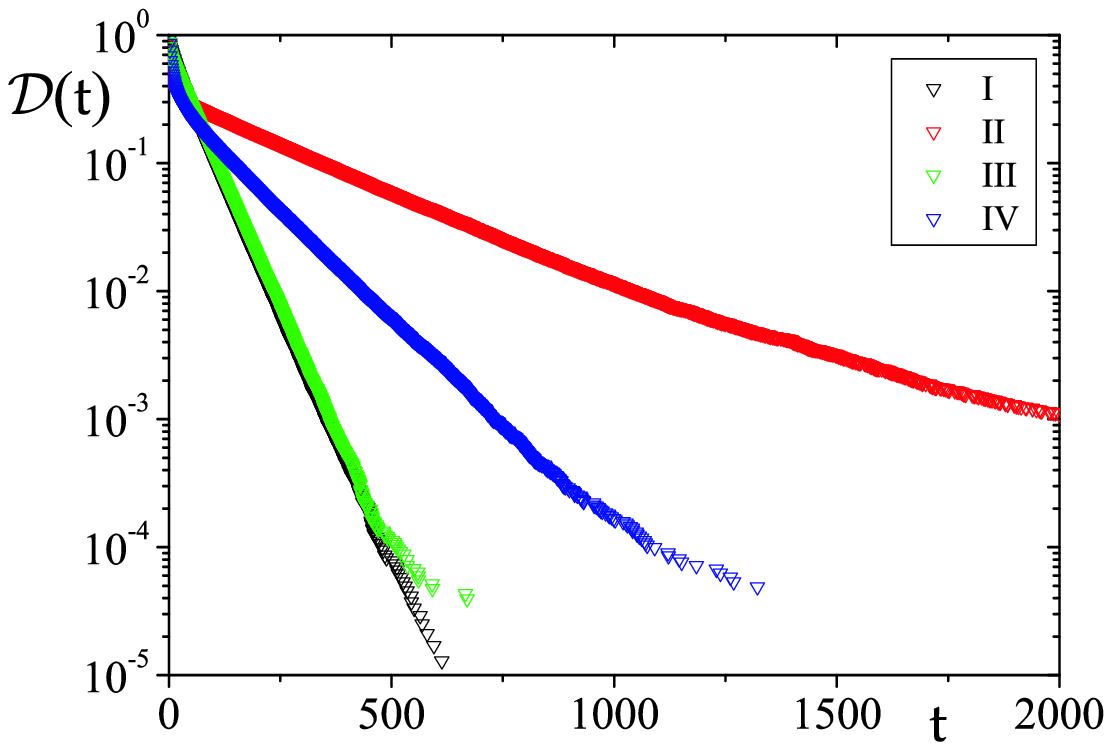} %
\includegraphics[angle=0,width=0.45\columnwidth]{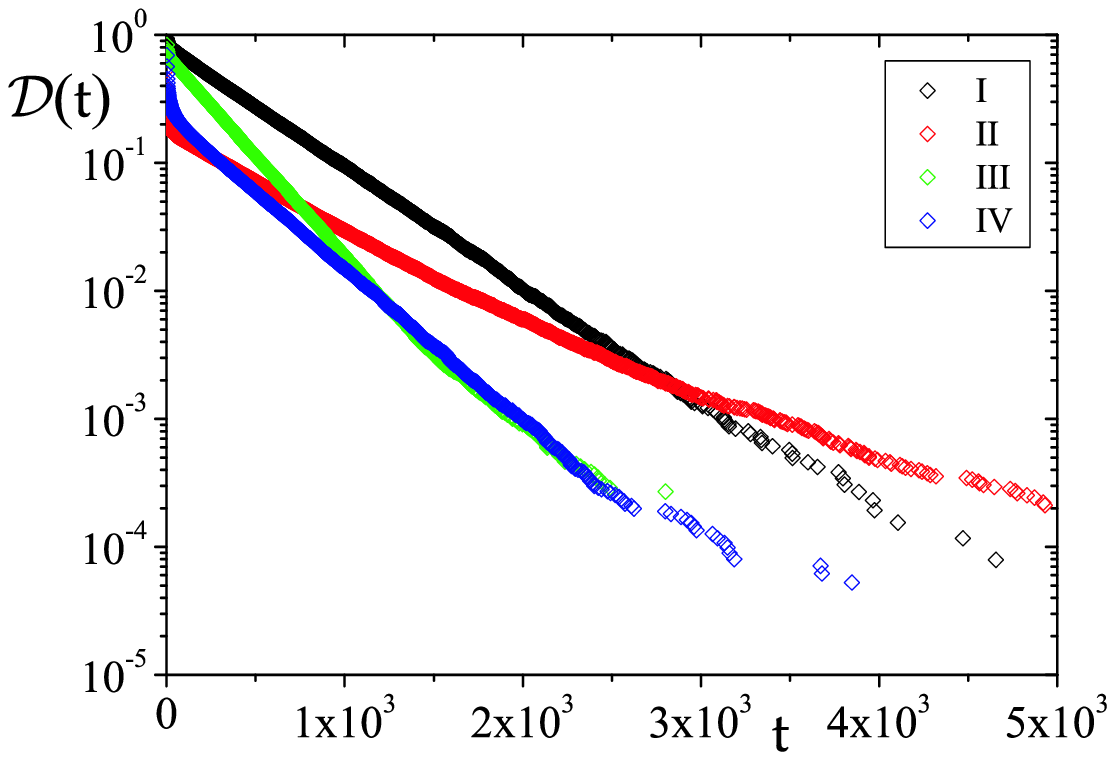} %
\includegraphics[angle=0,width=0.45\columnwidth]{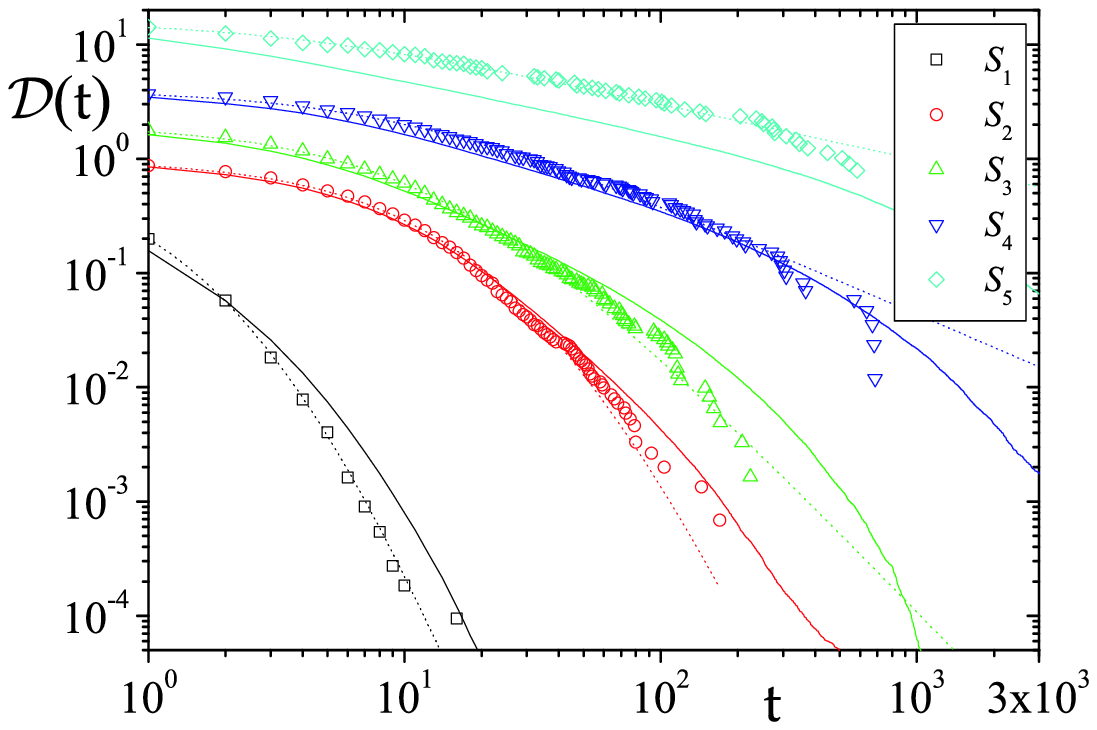}
\caption{Inverse cumulative distribution $\mathcal{D}\left( \mathfrak{t}%
\right) $ of the first-passage times{\protect\Large \ }\textit{vs}$\mathfrak{%
t}$ for the sets of parameters mentioned in the text. Upper panels: region $%
S_{1}$ (left)\ and $S_{2}$ (right); Middle panels: region $S_{3}$ (left) and
$S_{4}$ (right); Lower left panel: region $S_{5}$ (left). For the regions $%
S_{2}-S_{5}$, the exponential decay is evident. Lower right panel: Inverse
cumulative distribution $\mathcal{D}\left( \mathfrak{t}\right) $ of the
first-passage times{\protect\Large \ }\textit{vs} $\mathfrak{t}$ for squared
daily price fluctuations of $SP500$ (symbols) with the best fit represented
by the dotted lines. The full lines are obtained from the model with
parameters $q_{m}=q_{SP}$ and $b=b_{SP}$. The non-exponential behaviour is
clear in this latter case. The curves of the $S_{3}-S_{5}$ regions are
shifted by a factor of 2, 4, and 16, respectively.}
\label{fig-gi}
\end{figure}

Analysing the probability density functions we have verified that the
simplest expression which enables a numerical description of first-passage
inverse cumulative probability distribution, $\mathcal{D}\left( \mathfrak{t}
\right) $, is a linear composition of a asymptotic power-law (or a $\nu $%
-exponential) with a stretched exponential,
\begin{equation}
\mathcal{D}\left( \mathfrak{t}\right) =\epsilon \left[ 1+\left( \nu
-1\right) \frac{\mathfrak{t}}{\mathfrak{T}_{1}}\right] ^{\frac{1}{1-\nu }%
}+\left( 1-\epsilon \right) \exp \left[ -\left( \frac{\mathfrak{t}}{%
\mathfrak{T}_{2}}\right) ^{\phi }\right] .  \label{probtempo}
\end{equation}
Curves of some analysed cases are presented in Fig.~\ref{fig-gi} and fitting
parameters in Tab.~\ref{tab-gi}. The cases we present are: $I$-$\left\{
q_{m}=q_{SP},b=0.5\right\} $, $II$-$\left\{ q_{m}=1.25,b=b_{SP}\right\} $, $%
III$-$\left\{ q_{m}=1.25,b=0.5\right\} $, $IV $-$\left\{
q_{m}=1.5,b=0.875\right\} $.

From the Fig.~\ref{fig-gi} we have verified that, excepting region $S_{1}$,
all curves of $\mathcal{D}\left( \mathfrak{t}\right) $ exhibit a decay
closely exponential ($\nu =1$). For region $S_{1}$, as we increase the
non-Gaussianity of $p(z_{t})$, we have observed that both of the values of $%
\epsilon $ and $\nu $ approach one. Comparing the remaining curves, we have
verified that, for every region $S_{2}-S_{5}$, the set of parameters which
provides higher degree of non-Gaussianity has the larger characteristic
times $\mathfrak{T}$. Keeping the memory parameter $q_{m}$ constant, we have
observed that the higher $b$, the higher $\mathfrak{T}$ is. An inverse
dependence is found when we have fixed $b$ and let $q_{m}$ vary. In other
words, smaller values of $q_{m}$ (which enhance broader distributions) have
larger values of $\mathfrak{T} $. We have also verified that the
first-passage times are not Poisson distributed as it is straightforwardly
verified in $S_{2}$ plots. Looking at the values of squared daily index
fluctuations of $SP500$, we have verified that they present an asymptotic
power-law behaviour for $\mathcal{D}\left( \mathfrak{t}\right) $ ($\epsilon
=1$, $\nu >1$) (fitting parameters shown in Tab.~\ref{tab-gsp}). Comparing
the results of the model with empirical results from $SP500$ time series, we
have observed that the model provides an overall reasonable description of
first-passage times with curves almost superposing for $S_{2}-S_{4}$
regions. For $S_{1}$ and $S_{5}$ regions curves present similar exponents,
but different values of $\mathfrak{T}$. It is worth remembering that we can
improve the results by considering some characteristic time in $\mathcal{K}%
\left( t\right) $ equation. Taking into account the $\nu $ exponents
obtained for the adjustment of $SP500$ first-passage times, we verify that
larger and smaller $\nu $- exponents are quite different. Such a strong gap
invalidates, at a daily scale, the collapse (existence of a single exponent)
of $\mathcal{D}\left( \mathfrak{t}\right) $ curves proposed for
high-frequency data~\cite{stanley-returns}.

\subsection{Multiscaling properties}

Multiscaling has been the focus of several studies in the field of
complexity~\cite{mandelbrot-feder}, particularly regarding applications to
finance~\cite{mandelbrot-fractal}\cite{tiziana}\cite{eisler}. If in many
works multiscaling (multifractal) properties of price fluctuations are
presented, other studies have put those multiscaling properties into questioning~%
\cite{aparente-ms}. In this section, we analyse mainly multifractal
properties of $z_{t}$ and $z_{t}^{2}$. To this aim, we define a generic variable
\begin{equation}
Z_{t}^{\left( 2\right) }\left( T\right) \equiv
\sum_{i=1}^{T}z_{t+i-1}^{\left( 2\right) }.
\end{equation}
From it, we compute,
\begin{equation}
\Omega _{h}\left( T\right) \equiv \left\langle \left\vert Z_{t}^{\left(
2\right) }\left( T\right) -\left\langle Z_{t}^{\left( 2\right) }\left(
T\right) \right\rangle \right\vert ^{h}\right\rangle .
\end{equation}
If there is multiscaling, then the following scaling property is observed,
\begin{equation}
\Omega _{h}\left( T\right) \propto T^{\eta \left( h\right) }.
\end{equation}
The computation of $\Omega _{h}\left( T\right) $ has been made using the
well-known MF-DFA procedure \cite{mf-dfa}.

For the case of $z_{t}$, the multiscaling can be easily and analytically
explained. The multiscaling properties of a time series can emerge twyfold:
from memory and from non-Gaussianity. Since, by definition, the
heteroskedastic process we present is uncorrelated, then the only
contribution to multiscaling comes from the non-Gaussian character of
probability density functions. In this way, time series $\left\{
z_{t}\right\} $ is not a multifractal, but a \textit{bifractal} instead \cite%
{mf-dfa}. Therefore, the $\eta \left( h\right) $ curve is defined as
follows,
\begin{equation}
\eta \left( h\right) =\left\{
\begin{array}{ccc}
h\frac{q-1}{3-q} & \mathrm{for} & h<\frac{3-q}{q-1} \\
0 & \mathrm{for} & h>\frac{3-q}{q-1}%
\end{array}%
\right. .  \label{bifractal}
\end{equation}

\begin{figure}[h]
\centering
\includegraphics[angle=0,width=0.75\columnwidth]{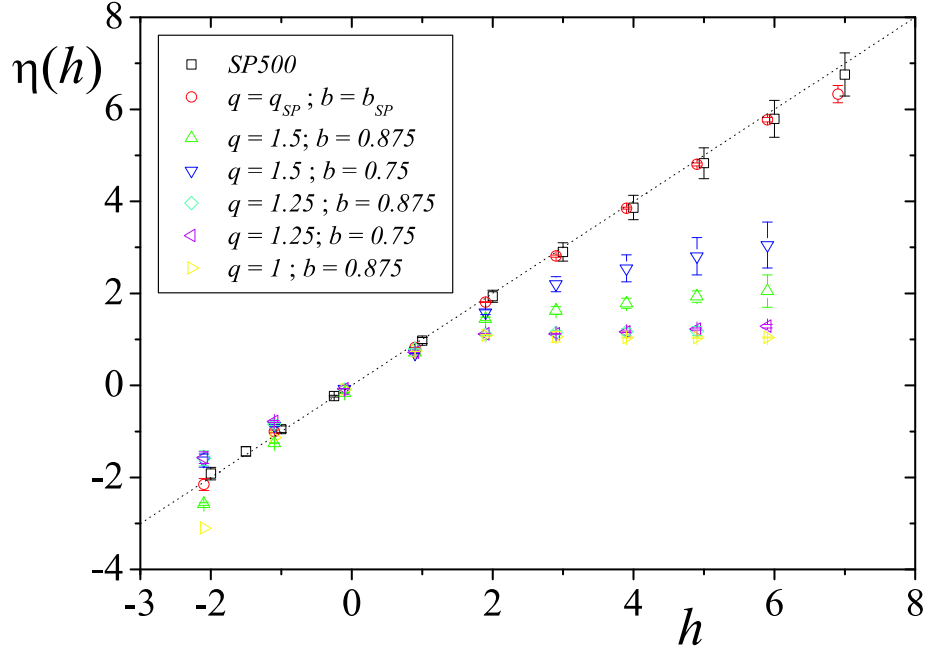}
\caption{Multiscaling exponents $\protect\eta $ \textit{vs} moments $h$ of
the values presented in the plot. The dashed line is $\protect\eta =h$.}
\label{fig-ms}
\end{figure}

With respect to $Z_{t}^{2}$ contrastive properties of $Z_{t}$ are found. The
results obtained from $SP500$ time series and the surrogate data have
enabled us to verify that the model is adequate to reproduce the scaling
properties of $Z_{t}^{2}$ which are basically linear according to $\eta
\left( h\right) =h$. For a constant $b$ value, we have observed that higher
correlations, introduced by increasing the value of $q_{m}$ turn
multiscaling properties weaker. In other words, $\eta _{z^{2}}\left(
h\right) $ approaches the straight line $\eta _{z^{2}}\left( h\right) =h$.
Similar qualitative results have been presented for traded value where
long-lasting correlations dominate specially for highly liquid equities~\cite%
{eisler}. Freezing memory parameter, $q_{m}$, we have verified similar
results, \textit{i.e.}, increasing the value of $b$, we increase the tails
in $z^{2}_{t}$ forcing the multiscaling curve to divert (even more) from the
straight line $\eta= h$. The same effect is obtained when memory is
shortened by reducing the value of $q_{m}$. By this we mean that, when
memory decays faster we have a detour from the straight line $\eta =h$ and
an approximation to a bifractal behaviour because of the asymptotic
power-law behaviour. This reflects the fact that the dynamical and
statistical properties of our system strongly depend on the ``force
relation'' between $b$ and $q_{m}$. Some results from which these
observations can be confirmed are shown in Fig.~\ref{fig-ms}.

\section{Asymmetric variance model~\label{assimetrico}}

In several systems it has been verified that the correlation function
between the observable and its instantaneous variance exhibits an
anticorrelation dependence. For example, this occurs in the case of
financial markets, when the correlation between past price fluctuations and
present volatilities is measured. The shape of the curve copes with the so called \textit{leverage effect}~%
\cite{haugen,bouchaud-prl}. This feature is intimately related to the risk
aversion phenomenon, \textit{i.e.}, falls in price turn the market more
volatile than price rises. In order to reproduce this characteristic we
introduce a small change in Eq. (\ref{zefect}), specifically,
\begin{equation}
\tilde{z}_{t-1}^{2}=\sum\limits _{i=t_{0}}^{t-1}\mathcal{K}\left(
i-t+1\right) \,\left[ z_{i}\left( 1-c\,z_{i}\right) \right] ^{2},
\label{zefectlev}
\end{equation}
where $c\geq 0$. It is worth stressing that this modification does not
introduce any skewness in the distribution $P\left(z\right) $ which is still
symmetric. It only acts on how positive and negative values of $z_{t}$, with
the same magnitude, influence $\tilde{z}_{t}^{2}$ by different amounts.

\begin{figure}[h]
\centering
\includegraphics[angle=0,width=0.75\columnwidth]{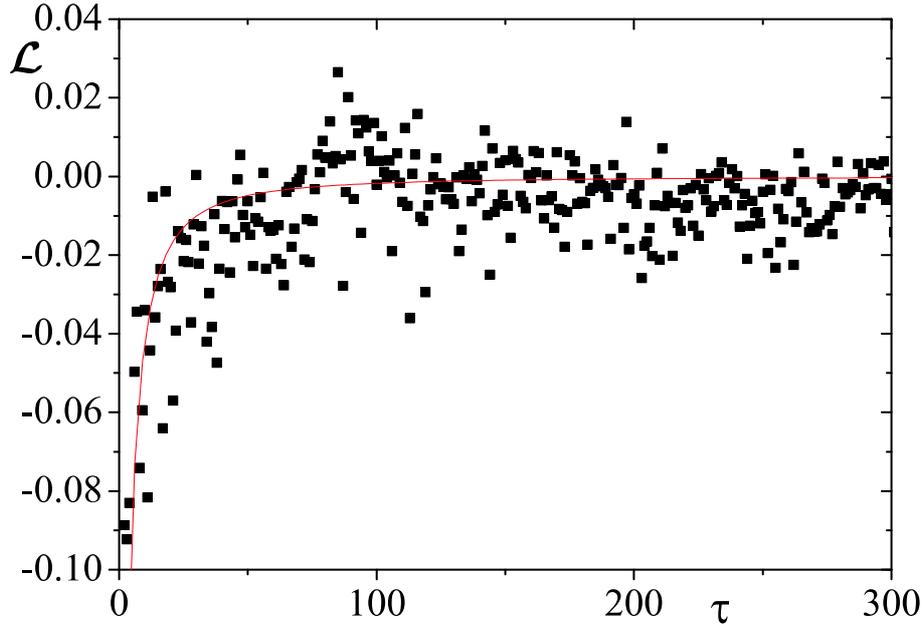}
\caption{The symbols represent leverage $\mathcal{L}$ \textit{vs} time lag $%
\protect\tau $ of a processes using Eq.~(\protect\ref{zefectlev}) with $%
q_{m}=1.65$, $b=0.95$ and $c=0.1$ obtained from a time series of $10^{6}$
elements. The red line is the best fit using Eq.~(\protect\ref%
{leverage-qarch}) with $\mathcal{L}\left( \protect\tau =0\right) =-0.39$ ($%
\protect\chi ^{2}=3\times 10^{-4}$).}
\label{fig-leverage}
\end{figure}

Using Eqs.~(\ref{vol-qarch})~and~(\ref{zefectlev}) in Eq.~(\ref{arch-def})
and expanding it up to first order we have
\begin{equation}
z_{t}=\left\{ \sqrt{a}+\frac{b}{2}\sum\limits _{i=t_{0}}^{t-1}\mathcal{K}%
\left( i-t+1\right) \left[ z_{i}^{2}-2\,c\,z_{i}^{3}+c^{2}z_{i}^{4}\right]
+\ldots \right\} \,\omega _{t}.  \label{z-leverage}
\end{equation}
Computing $z_{t}\,z_{t+\tau }^{2}$ the only terms which do not vanish after
averaging are
\begin{equation}
\mathcal{T}_{1}=-2\sqrt{a}\,b\,c\sum\limits _{i=t_{0}}^{t-1+\tau }\mathcal{K}%
\left( i-\left( t-1+\tau \right) \right) \,z_{i}^{3}\,\omega _{t}\,\omega
_{t+\tau }^{2},
\end{equation}
and
\begin{equation}
\begin{array}{cc}
\mathcal{T}_{2}= & -2\,b\,c^{3}\sum\limits_{i=t_{0}}^{t-1+\tau
}\sum\limits_{j=t_{0}}^{t-1}\mathcal{K}\left( i-\left( t-1+\tau \right)
\right) \mathcal{K}\left( j-t+1\right) \times \\
&  \\
& z_{i}^{3}\,z_{j}^{4}\,\omega _{t}\,\omega _{t+\tau }^{2}.%
\end{array}
\,
\end{equation}
Performing averages and considering stationarity we have in the continuous
limit,
\begin{equation}
\begin{array}{cc}
\left\langle \mathcal{T}_{1}\right\rangle = & --6\sqrt{a}\,b\,c\int_{0}^{t+%
\tau }\mathcal{K}\left( x-\left( t+\tau \right) \right) \times \\
&  \\
& \left\langle z_{t}^{2}\right\rangle \,\left\langle \sigma
_{t}\right\rangle \,\left\langle \omega _{t}^{2}\right\rangle \,\delta
\left( x-t\right) \,dx,%
\end{array}
\,
\end{equation}
which gives,
\begin{equation}
\left\langle \mathcal{T}_{1}\right\rangle \sim -\exp _{q_{m}}\left[ -\tau %
\right] .
\end{equation}
\begin{equation}
\begin{array}{cc}
\mathcal{T}_{2}= & -2\,b\,c^{3}\left\{ \int_{0}^{t}\mathcal{K}\left(
x-\left( t+\tau \right) \right) \mathcal{K}\left( x-t\right)
\,z_{x}^{7}\,\omega _{t}\,\omega _{t+\tau }^{2}\,dx+\right. \\
&  \\
& 2\int_{0}^{t}\int_{x}^{t}\mathcal{K}\left( x-\left( t+\tau \right) \right)
\mathcal{K}\left( y-t\right) z_{x}^{3}z_{y}^{4}\,\omega _{t}\,\omega
_{t+\tau }^{2}\,dx\,dy+ \\
&  \\
& \left. \int_{t}^{t+\tau }\int_{0}^{t}\mathcal{K}\left( x-\left( t+\tau
\right) \right) \mathcal{K}\left( y-t\right) z_{x}^{3}z_{y}^{4}\,\omega
_{t}\,\omega _{t+\tau }^{2}\,dx\,dy\right\} .%
\end{array}
\label{t2}
\end{equation}%
Averaging, only the first integral has a non-null contribution yielding,
\begin{equation}
\left\langle \mathcal{T}_{2}\right\rangle =-2\,b\,c^{3}\left( 2-q_{m}\right)
^{2}\exp _{q_{m}}\left[ -\tau \right] \left\langle z_{t}^{6}\right\rangle
\left\langle \sigma _{t}\right\rangle \left\langle \omega
_{t}^{2}\right\rangle .
\end{equation}
It is not hard to show that\footnote{%
In order to keep the formulae as simple as possible we use the following
expressions the discrete notation $z_{t}$. Formally, it should be read as $%
z\left( t\right) $ since we are dealing with a continuous approach.}
\begin{equation}
\begin{array}{cc}
\left\langle z_{x}^{3}z_{y}^{4}\,\omega _{t}\right\rangle = & 5\,\,\delta
\left( x-t\right) \left\langle \sigma _{t}\right\rangle \left\langle
z_{t}^{2}\right\rangle \left\langle z_{t}^{4}\right\rangle + \\
&  \\
& 24\,\delta \left( y-t\right) \delta \left( x-y\right) \left\langle \sigma
_{t}\right\rangle \left\langle \sigma _{t}^{2}\right\rangle \left\langle
z_{t}^{2}\right\rangle ^{2}+ \\
&  \\
& 72\,\delta \left( x-t\right) \left[ \delta \left( x-y\right) \right]
^{2}\left\langle \sigma _{t}\right\rangle \left\langle \sigma
_{t}^{2}\right\rangle \left\langle z_{t}^{2}\right\rangle + \\
&  \\
& 12\,\delta \left( x-y\right) \left\langle \sigma _{t}\right\rangle
\left\langle z_{x}^{2}z_{y}^{3}\omega _{t}\right\rangle + \\
&  \\
& 12\,\delta \left( y-t\right) \delta \left( x-y\right) \left\langle \sigma
_{t}\right\rangle ^{2}\left\langle z_{x}^{2}z_{y}^{2}\right\rangle .%
\end{array}
\label{wilk}
\end{equation}
Inserting Eq.~(\ref{wilk}) in Eq.~(\ref{t2}) the last two integrals give
zero. Therefore, in the first approximation, the leverage function,
\begin{equation}
\mathcal{L}\equiv \frac{\left\langle z_{t}\,z_{t+\tau }^{2}\right\rangle }{%
\left\langle z_{t}^{2}\right\rangle ^{2}},  \label{leverage}
\end{equation}
goes as,
\begin{equation}
\mathcal{L}\sim \left\langle \mathcal{T}_{1}\right\rangle +\left\langle
\mathcal{T}_{2}\right\rangle \sim -\exp _{q_{m}}\left[ -\tau \right] .
\label{leverage-qarch}
\end{equation}
As it can be seen from Fig.~\ref{fig-leverage}, the approximation provides a
satisfactory approximation of the numerical results. A precise description
can obviously be obtained by considering higher-order (slowly decaying)
terms which are obtained through a quite tedious computation that follows
exactly the same lines we have just introduced.

\begin{figure}[h]
\centering
\includegraphics[angle=0,width=0.75\columnwidth]{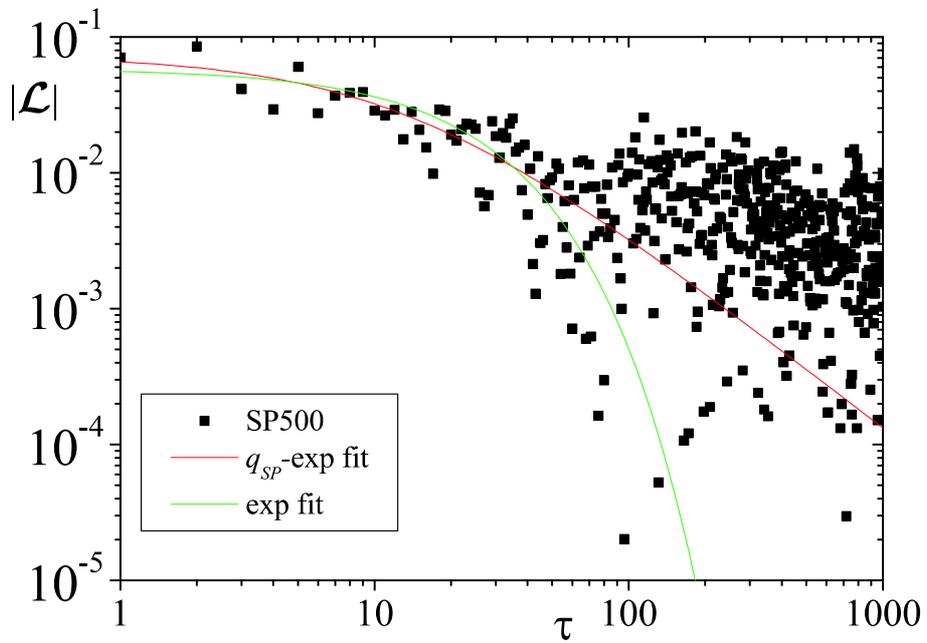}
\caption{The symbols represent the absolute value of leverage $\mathcal{L}$
\textit{vs} time lag $\protect\tau $ for the daily index fluctuations of $%
SP500$ spanning the period mentioned above. The red line represents the
fitting for a $q$-exponential function with $L_{0}=0.07$, $q=q_{SP}$ and $%
T=9 $, and the green line a exponential fitting with $L_{0}=0.06$ and $T=20$
(fitting error values in the text).}
\label{fig-leveragesp500}
\end{figure}

The result above is in apparent contradiction with previous work in which an
exponential dependence with $\tau $ is defended in lieu of an asymptotic
power-law dependence. Nevertheless, in Fig.~\ref{fig-leveragesp500} we show
the leverage function computed from the $SP500$ and numerical adjustments
with function,
\begin{equation}
L\left( \tau \right) =-L_{0}\exp _{q_{m}}\left[ -\frac{\tau }{T}\right] ,
\end{equation}
with $q_{m}=q_{SP}$ and $q_{m}=1$. Computing the adjustment error, $\chi
^{2} $ and $R^{2}$, we verify that both approaches present similar values
for the numerical adjustment with $q=q_{SP}$ being scanty better. For $%
q=q_{SP}$, we have obtained $\chi ^{2}=4\times 10^{-5}$ and $R^{2}=0.47$ and
for the exponential adjustment $\chi ^{2}=5\times 10^{-5}$ and $R^{2}=0.44$.
From these results, we can affirm that our proposal is, at least, as good as
the exponential decay scenario firstly introduced in Ref.~\cite{bouchaud-prl}%
. It is also worth noting that, although this variation is asymmetric
concerning the effects of the sign of variables $z_{t}$ on the evaluation of
$\sigma _{t}^{2}$, the model is ineffective in reproducing the skewness of
the distribution of price fluctuations. This owes to the fact that $\omega
_{t}$ used up to now is symmetric thus, also annulling the asymmetry introduced
by the Eq.~(\ref{zefectlev}). It is worth stressing that moments $%
\left\langle z_{t}^{n}\right\rangle $, for $n$ even, in this section are
obviously different from the values presented in preceding sections.

\section{Final remarks~\label{final-remarks}}

In this manuscript we have introduced further insight into a heteroskedastic
process enclosed in the class of fractionally integrated $ARCH$ processes.
This process is characterised by a memory of past values of the squared
variable which decays according to an $q_{m}$-exponential. Despite of the
fact that we were unable to provide an analytical proof, prevailing statistical testing
has shown that $q$-Gaussian distributions properly describe the probability
density function of the generated stochastic variable. Based on this fact,
we have determined a form of the instantaneous variance, $\sigma ^{2}$,
probability density function which has yielded a inverse Gamma distribution
like it happens in superstatistical models giving $q$-Gaussians as long-term distributions. Moreover, we
have computed the first term of the correlation function, $%
C_{\tau }\left( z_{t}^{2}\right) $, which corresponds to a $q_{c}$%
-exponential. An analytical relation between $q_{m}$ and $q_{c}$ is
presented. From these results, we are able to state that this dynamical
system can actually be described within non-extensive statistical mechanics
(NESM) framework by a triplet of $q$ values \cite{q-tripleto}. As a matter
of fact, this process presents all the elements to be characterised, in $%
z_{t}^{2}$ variable, as a NESM process. Explicitly, besides presenting
asymptotic power-law distributions which maximise non-additive entropy $%
S_{q} $ (as $z_{t}$), it has a slow decaying ($q_{c}$-exponential)
auto-correlation function, and it exhibits multiscaling properties. Such
properties have been advocated as primary features of
systems that should be studied within NESM framework (for related literature
see~\cite{livros-NESM}) for a long time. Furthermore, we have verified that, for a
sufficiently high level of memory the model presents a non-exponential
distribution of first-passage times and strong levels of dependence measured
from a generalisation of Kullback-Liebler mutual information. Though they
have been applied in several other areas, in view of the fact that
hetoskedastic processes have been introduced in a financial context, we have
tested the model against daily index fluctuations of $SP500$. The results
firstly presented, together with the results of this manuscript, show that
this model, despite of its simplicity (\textit{it only has two parameters}),
is able to reproduce the most relevant and important properties, namely the
probability density functions, the Hurst exponent, the autocorrelation
functions, multiscaling, and first-passage time distribution (in a less good
extend compared to previous). It is our belief that the same occurs with
other time series presenting similar characteristics. Moreover, we have
studied how the extension of the memory (tuned by $q_{m}$) and its weight,
or alternatively, the weight of fluctuations in $\sigma _{t}^{2}$ (adjusted
by $b$) have on the quantities. Still in the context of financial time
series, we have introduced a slight modification which allows the
reproduction of the leverage effect. Under these circumstances, we propose
that the leverage is not described for an exponential function, but for a $%
q_{m}$-exponential function instead. When statistically tested, this
proposal has emerged as good as the exponential description. It is
well-known that many distributions obtained from complex systems present
skweness which this model has not been able to capture because of the
symmetrical nature of noise $\omega $ distribution. The use of other types
of noise $\omega $, jointly with the modification presented in Sec.~\ref%
{assimetrico} might give rise to even more precise modelling.

\subsection*{Acknowledgements}

SMDQ acknowledges C. Tsallis for several comments and discussions on matters
related to this manuscript. This work benefited firstly from financial
support from FCT/MCES (Portuguese agency)\ and infrastructural support from
PRONEX/MCT (Brazilian agency) and in its later part from financial support
from Marie Curie Fellowship Programme (European Union).


\begin{table}[tbp]
\caption{Table of the fitting parameters of region $S_{1}$using Eq. (\ref{probtempo})}
\label{tab-gi}%
\begin{tabular}{ccccc}
\hline
& $I$ & $II$ & $III$ & $IV$ \\ \hline\hline
$\epsilon $ & $0.388$ & $0.004$ & $0.414$ & $0.334$ \\
$\nu $ & $1.09$ & $1$ & $1.17$ & $1.17$ \\
$\mathfrak{T}_{1}$ & $0.648$ & $2.23$ & $0.541$ & $0.951$ \\
$\phi $ & $0.894$ & $0.394$ & $0.792$ & $0.762$ \\
$\mathfrak{T}_{2}$ & $0.867$ & $0.001$ & $0.757$ & $0.27$ \\
$\chi ^{2}$ & $5.1\times 10^{-9}$ & $9.2\times 10^{-7}$ & $2.6\times 10^{-6}$
& $1.9\times 10^{-6}$ \\ \hline
\end{tabular}%
\end{table}

\begin{table}[tbp]
\caption{Table of the fitting parameters of dily index fluctuation of $SP500$
region using Eq.(\protect\ref{probtempo}).}
\label{tab-gsp}%
\begin{tabular}{cccccc}
\hline
& $S_{1}$ & $S_{2}$ & $S_{3}$ & $S_{4}$ & $S_{5}$ \\ \hline\hline
$\epsilon $ & $1$ & $1$ & $1$ & $1$ & $1$ \\
$\nu $ & $1.17$ & $1.21$ & $1.43$ & $2.03$ & $3.03$ \\
$\mathfrak{T}_{1}$ & $1.85$ & $0.14$ & $0.157$ & $0.101$ & $0.143$ \\
$R^{2}$ & $0.999$ & $0.999$ & $0.998$ & $0.998$ & $0.993$ \\
$\chi ^{2}$ & $5.4\times 10^{-7}$ & $5\times 10^{-5}$ & $5\times 10^{-5}$ & $%
1\times 10^{-4}$ & $3\times 10^{-4}$ \\ \hline
\end{tabular}%
\end{table}

\end{document}